\documentclass[a4paper,10pt]{article}
\usepackage[english]{babel}
\usepackage[utf8]{inputenc}
\usepackage{color}
\usepackage{amsmath,amsfonts}
\usepackage{amssymb}
\usepackage{caption,subcaption,units}
\usepackage{graphicx}
\color{black}

\begin{document}

\title{Periodic and solitary wave solutions of the long wave-short wave Yajima-Oikawa-Newell model}

\author{%%%% Author details
Marcos Caso-Huerta \\ Department of Mathematics, Physics and Electrical Engineering \\ Northumbria University, Newcastle upon Tyne, UK \\ Email: marcos.huerta@northumbria.ac.uk \\ {} \\
Antonio Degasperis \\ Dipartimento di Fisica \\ ``Sapienza'' Universit\`a di Roma, Rome, Italy \\ Email: antonio.degasperis@uniroma1.it \\ {} \\
Priscila Leal da Silva \\ Department of Mathematical Sciences, School of Science \\ Loughborough University, Loughborough, UK \\ Email: p.leal-da-silva@lboro.ac.uk  \\ {} \\
Sara Lombardo \\ Department of Mathematical Sciences, School of Science \\ Loughborough University, Loughborough, UK \\ Email: s.lombardo@lboro.ac.uk \\ {} \\
Matteo Sommacal \\ Department of Mathematics, Physics and Electrical Engineering \\ Northumbria University, Newcastle upon Tyne, UK \\ Email: matteo.sommacal@northumbria.ac.uk
}

\date{}

\maketitle

\begin{abstract}
Models describing long wave-short wave resonant interactions have many physical applications from fluid dynamics to plasma physics. We consider here the Yajima-Oikawa-Newell (YON) model, which has been recently introduced combining the interaction terms of two long wave-short wave, integrable models, one proposed by Yajima-Oikawa, and the other one by Newell.

The new YON model contains two arbitrary coupling constants and it is still integrable  -- in the sense of possessing a Lax pair -- for any values of these coupling constants. It reduces to the Yajima-Oikawa or the Newell systems for special choices of these two parameters.
We construct families of periodic and solitary wave solutions, which display the generation of very long waves. We also compute the explicit expression of a number of conservation laws.
\end{abstract}

%%%%%%%%%%%%%%%%%%%%%%%%%%%
\section{Introduction}\label{sec:intro}
The study of wave propagation poses quite a number of challenging different mathematical and computational problems \cite{Whitham}. %In order to cope with the large number of different waves, few concepts have been singled out.

Wave motion of continuous media is generally represented by solutions of one or more partial differential equations (PDEs). Typically, homogeneous wave propagation equations have a linear part which is characterised by a dispersion law, and a nonlinear one which is responsible for self and/or cross interaction. While the linear part can be treated by decomposition into Fourier harmonics, the nonlinear part -- even if its dependence on the fields is analytic -- is generically treatable only by numerical methods. These two parts, in terms of solvability, recombine in cooperation with each other for a small set of very special wave equations, the so-called \emph{integrable} ones, by allowing for a sort of nonlinear Fourier-like analysis \cite{AKNS,CalDe}.

We have in mind (and consider below) only waves propagating in a one-dimensional space, even if integrable wave equations in higher dimension are known too. Since their first discovery more than half a century ago, the firmament of integrable wave equations has been continuously growing. From the very beginning, water waves have played a pivotal role in this research, with the discovery of the integrability of the Boussinesq equation, the Korteweg-de Vries (KdV) equation and the nonlinear Schr\"{o}dinger (NLS) equation, which may serve as approximate models of waves travelling on a water free surface subject to various physical conditions.

The existence of solitons is the most celebrated and effective prediction from the study of water wave equations of integrable type. Subsequent research has pointed out that discovering integrable wave equations is not only due to a lucky strike in the process of approximating complicated PDEs of physical significance. In fact, integrable models can be obtained by means of a perturbation approach, the so-called multiscale method (see \cite{Zakharov-Kuznetsov1986,Calogero1991,Degasperis2009}), when applied to a given known nonlinear wave equation. Indeed, given a physically relevant PDE, by appropriately rescaling both wave amplitudes and space-time coordinates via the introduction of a small parameter $\epsilon$, and by expanding in powers of $\epsilon$, one ends up with a different PDE which models the amplitude modulation in the rescaled space-time coordinates. The point is that this process preserves the integrability property of the original wave equation and thus it yields a (possibly) \emph{new} integrable wave equation. The rationale behind this method is based on physical arguments: one considers the nonlinear terms as a perturbation of known Fourier-like solutions (harmonics) of the linear dispersive part. Consequently, approximate solutions of the original PDE look like superpositions of exponentials as, for instance, $\sum_j\psi_j e^{i\,j\,(kx-\omega t)}$, whose amplitudes $\psi_j$ depend on rescaled coordinates only. The parameters $k$ and $\omega$ are the wave number and frequency on the dispersion curve $\omega=\omega(k)$. If the superposition of harmonics contains more wave-numbers, this multiscale method in general yields a system of coupled PDEs which models wave-wave interactions.

A necessary condition for this to happen is the (weak) resonance relation:
$$
\omega(k_1+k_2 +\cdots+k_j)=\omega(k_1)+\omega(k_2)+\cdots+\omega(k_j),\quad j>1\,.
$$
Several models of resonant interaction have been derived in this way (\textit{e.g.}, see \cite{Calogero1991,CDJ12000,Degasperis2009}) to investigate their main features by means of known integrability techniques, provided such models were selected as sufficiently ``close'' to a wave equation of specific physical interest.

This perturbative approach is also appropriate to investigate the resonant coupling of two quasi-monochromatic waves, one with very long wave-length, say with wave number $k_L\approx 0$, and the second one with a much shorter wave-length, say with wave number $k_S\gg k_L$. As originally pointed out in \cite{Benney1977}, this interaction can be understood as a resonant triad $k_1$, $k_2$, $k_3$, namely
$$
k_1+k_2=k_3\,,\quad
\omega(k_1)+\omega(k_2)=\omega(k_3)\,,
$$
with
$$
k_1=k_L\,,\qquad
k_2=k_S-(1/2)k_L\,,\qquad
k_3=k_S+(1/2)k_L\,.
$$
Indeed, if the long wave is sufficiently long, say $k_L\rightarrow 0$, and the dispersion function $\omega(k)$ is analytic at $k=0$, this condition is equivalent to the stronger condition that the long wave and the short wave have the same group velocity, $v_L=v_S$.

The search for integrable PDEs which reasonably model phenomena due to long wave-short wave (L-S) interaction started in the early years of the soliton era in fluid dynamics, plasma physics and optics. The Yajima-Oikawa (YO) system \cite{YO1976}
\begin{equation}
\label{YO}
i S_t +S_{xx} - LS=0\,,\quad L_t=2(|S|^2)_x
\end{equation}
was first derived in the one-way wave approximation in plasma. Here and in  the following $S$ and  $L$ are the complex, and, respectively, real amplitudes of the short and long waves. This system shows up also via multiscale technique \cite{CDJ12000} in multiple ways. In fact, it proves to be a multiscale reduction of an integrable equation of interest in water waves, namely the Boussinesq equation \cite{Calogero1991}.
A second and alternative integrable L-S wave system has been proposed by Newell \cite{Newell1978}. This one reads
\begin{equation}
\label{N}
i S_t +S_{xx} +( i L_x + L^2 -2 \sigma |S|^2) S=0\,,\quad
L_t=2 \sigma (|S|^2)_x \,,\quad \sigma^2=1\,,
\end{equation}
where, in addition to a long wave-short wave coupling, the short wave has the same self-interaction as the NLS equation, which may be both defocusing $(\sigma=1)$ or focusing $(\sigma=-1)$ according to the sign $\sigma$. We will refer to the system (\ref{N}) as to the N equation. However, the way to obtain this system as a multiscale reduction of an integrable equation does not seem to be known.

In \cite{CDLS2021}, it has been shown that the YO and N equations, (\ref{YO})  and (\ref{N}), need not to be separately treated to investigate the short wave-long wave interaction. Indeed, these two different model equations, remarkably enough, can be combined in just one system, which we refer to as YON model, that is itself integrable for \emph{any} real value of the two arbitrary parameters $\alpha$ and $\beta$, namely
\begin{equation}
\label{YON}
i S_t +S_{xx} +\left(i\alpha L_x+\alpha^2 L^2-\beta L -2\alpha |S|^2 \right) S=0\,,
\quad L_t=2 (|S|^2)_x \,.
\end{equation}
This system coincides with the YO equation (\ref{YO}) for $\alpha=0$, $\beta=1$, while it reads as the N equation (\ref{N}) by setting $\alpha=\sigma$, $\beta=0$ and by substituting the field $L$ with $\sigma\,L$. This unifying result provides a greater flexibility in modelling the resonant interaction of long and short waves, and allows to construct and analyse in one go special solutions of the two models, (\ref{YO}) and (\ref{N}). On the mathematical side, we note that the YON model (\ref{YON}) turns out to be a reduction of a larger system of four coupled PDEs \cite{Wright2006}.
%{\color{blue}It is worth observing that the YON system can be derived from system (53) in \cite{Wright2006}, a fact of which we became aware after the publication of \cite{CDLS2021}.}

The preliminary step to inquire on the application of the YON system to specific physical wave phenomena is the compatibility of the L-S resonance condition (see the triad resonance above) with the dispersion law (for an instance of such analysis in optics, see \cite{CT2008}). Having in mind a fluid dynamical context, we recall a few elementary facts to show that water waves on the free surface of a two-dimensional rectangular container require that gravity be contrasted by surface tension \cite{DR1977}. Indeed, this effect allows the resonance, which, however -- at least in geophysical applications -- is far from being of experimental relevance, see \cite{Lannes2013}. As it happens, if the surface tension is neglected, the dispersion law
$$
\omega^2 = g\,k\,\tanh(h\,k)\,,
$$
where $g$ is the gravity acceleration and $h$ is the depth of the flat bottom, does not allow for the L-S resonance since the group velocity $\mathrm{d}\omega(k)/\mathrm{d}k$ is monotonically decreasing in the entire range $0<k<+\infty$. Surface tension, if the wave length is sufficiently short, may contrast gravity and change the dispersion law into
$$
\omega^2 = g\,k\,\tanh(h\,k)\,\left[1+\frac{n\,k^2}{g\,\rho}\right]\,,
$$
where $n =0.074\,\mathrm{N}/\mathrm{m}$ is the surface tension constant and $\rho$ is the water density. Although this change of the frequency dispersion formally allows for the S-L resonance, the value of $k_S$ which satisfies the strong resonance condition $v_L=v_S$ strongly depends on the depth $h$. For instance, a short wave of about $1\,\mathrm{cm}$ length requires a flat bottom of approximately $1\,\mathrm{cm}$ depth. For stratified fluids, see for instance \cite{Grimshaw1977,KR1981}.
%%%%%%%
 %{KR1981} is DJORDJEVIC, V. D. & REDEKOPP, L. G. (1977) On two-dimensional packets of capillary-gravity waves.J. Fluid Mech., 79, 703Ã714
 %%%%%%%%%

Despite such resonant conditions lead to small effects in capillarity-gravity wave propagation, we deem it of interest to investigate the YON model because of its potential applicability while being integrable. Indeed it has been shown \cite{CDLS2021} that there exists a Lax pair of equations for an auxiliary function $\Psi(x,t,\lambda)$ (where $\lambda$ is a complex number called spectral parameter or spectral variable)
\begin{equation}
\label{laxpair}
\Psi_x=X\,\Psi\,, \quad \Psi_t=T\,\Psi\,\,,
\end{equation}
whose compatibility condition
\begin{equation}
\label{compat}
X_t - T_x + [X\,,\,T] =0
\end{equation}
is equivalent to the YON system (\ref{YON}). Here $X(x,t,\lambda)$ and $T(x,t,\lambda)$ are complex matrix-valued functions depending on the field variables $S$ and $L$, and also polynomially on the spectral variable $\lambda$,
\begin{equation}
\label{XTpair}
%X(\lambda)=i \lambda \Sigma + Q\,,\quad T(\lambda)= (i \lambda)^2 A + i\lambda B + C\,,
X=i \lambda\, X_1 + X_0\,,\quad T= (i \lambda)^2 T_2 + i\lambda\, T_1 + T_0\,,
\end{equation}
%where $\Sigma$ denotes the constant, traceless, diagonal matrix
where $X_1$ and $T_2$  denote constant, traceless, diagonal {matrices} %MDPI: subequation is not allowed, please check if you insist to keep, please check full text.

\begin{subequations}
\begin{equation}
\label{constantmatrix}
X_1=%\textrm{diag}(1 \,,\, 0\,,\, -1 )=
\left(\begin{array}{ccc} 1 & 0 & 0 \\  0 & 0 &  0  \\ 0 & 0  & -1 \end{array} \right)\,,\quad
T_2=\frac{i}{3}\left (\begin{array}{ccc} 1 & 0 & 0 \\ 0 & -2 & 0  \\0 & 0 & 1 \end{array} \right )\,,
\end{equation}
and the matrices $X_0$, $T_1$ and $T_0$ have the form
\begin{equation}
\label{X0T1}
X_0=\left (\begin{array}{ccc} 0 & S & iL \\ \alpha  S^\ast & 0 &   S^\ast  \\ i\alpha^2L-i\beta & \alpha S & 0 \end{array} \right )\,,\quad
T_1= \left (\begin{array}{ccc} 0 & i S & 0 \\ i \alpha S^\ast & 0 & -i S^\ast  \\ 0 & -i\alpha S & 0 \end{array} \right )\,,
\end{equation}
%\begin{equation}
%\label{AB}
%%A=\frac{i}{3}\left (\begin{array}{ccc} 1 & 0 & 0 \\ 0 & -2 & 0  \\0 & 0 & 1 \end{array} \right )\,,
%\quad T_1= \left (\begin{array}{ccc} 0 & i S & 0 \\ i \alpha S^\ast & 0 & -i S^\ast  \\ 0 & -i\alpha S & 0 \end{array} \right )\,,
%\end{equation}
\begin{equation}
\label{T0}
T_0= \left (\begin{array}{ccc} -i\alpha |S|^2 & -\alpha LS +iS_x & i|S|^2 \\ - \alpha^2 LS^\ast +\beta  S^\ast-i \alpha S^\ast_x  & 2i \alpha |S|^2 &  -\alpha LS^\ast -i S^\ast_x \\ i \alpha^2 |S|^2& -\alpha^2 LS +\beta S+i\alpha S_x & -i\alpha |S|^2 \end{array} \right )\,.
\end{equation}
\end{subequations}

However, in the following, we will not make use of the Lax pair (\ref{laxpair}), and we refer the interested reader to \cite{CDLS2021}, where further details on the integrable character of the system can be found.
%%%

%{\color{blue}
%We end this section by observing that system \eqref{YON} is trivially invariant with respect to translations in space and time, as well as to rotations %around the origin in the $S$-plane. Moreover, the YON system is invariant under the general transformation
%\[
%(x,\,t,\,S,\,L)\to \Big(\varepsilon^{-1}x,\,\varepsilon^{-2}t,\,\varepsilon\,\exp{\Big[i\frac{(\varepsilon^2 -1)\beta^2 %t}{4\alpha^2}\Big]}\,\,S,\,\varepsilon L-\frac{(\varepsilon-1)\beta}{2\alpha^2}\Big)
%\]
%where $\varepsilon\neq0$ is an arbitrary parameter. Note that the limit $\alpha\to 0$ of the above transformation can be obtained by mapping $\varepsilon %\to \exp{(-2\alpha^2\varepsilon)}$ first.
%If $\alpha=0$, we have that \eqref{YON} is invariant under the additional scaling:
%\[
%(x,\,t,\,S,\,L)\to (e^{2\varepsilon}x,\,e^{4\varepsilon}t,\,e^{-3\varepsilon}S,\,e^{-4\varepsilon}L) .
%\]
%%%%
%In the next section we discuss several families of periodic and traveling wave solutions of the YON system (\ref{YON}), including dark and bright solitons, %as well as rational solitons.
%%In Section \ref{sec:sym} we derive the Lie point symmetries possessed by the YON system.
%Finally, in Section \ref{sec:cl}, using a set of multipliers, we construct and exhibit a family of conservation laws, out of the infinity of conservations %laws enjoyed by the YON system as a result of its Lax-integrability.
%}

In the next section we discuss several families of periodic and traveling wave solutions of the YON system (\ref{YON}), including dark and bright solitons, as well as rational solitons. Finally, in Section \ref{sec:cl}, using a set of multipliers, we construct and exhibit a family of conservation  laws. To this purpose, it is instrumental the use of symmetry transformations of our system (\ref{YON}). Thus we end this section by observing that the YON system (\ref{YON}) is trivially invariant with respect to translations in space and time, as well as to rotations around the origin in the $S$-plane. Moreover, the YON system is invariant under the general transformation
\[
\left(x,\,t,\,S,\,L\right)\to \Big(\varepsilon^{-1}x,\,\varepsilon^{-2}t,\,\varepsilon\,\exp{\Big[i\frac{(\varepsilon^2 -1)\beta^2 t}{4\alpha^2}\Big]}\,\,S,\,\varepsilon L-\frac{(\varepsilon-1)\beta}{2\alpha^2}\Big)
\]
where $\varepsilon\neq0$ is an arbitrary parameter. Note that the limit $\alpha\to 0$ of the above transformation can be obtained by mapping $\varepsilon \to \exp{(-2\alpha^2\varepsilon)}$ first.
If $\alpha=0$, we have that \eqref{YON}, namely the YO system (\ref{YO}), is invariant under the additional scaling:
\[
\left(x,\,t,\,S,\,L\right)\to
%\left(e^{2\varepsilon}x,\,e^{4\varepsilon}t,\,e^{-3\varepsilon}S,\,e^{-4\varepsilon}L\right)\,,
\left(\varepsilon^2\,x,\, \varepsilon^4\,t,\, \varepsilon^{-3}\,S,\, \varepsilon^{-4}\,L\right)\,,
\]
while, if $\beta=0$, say if \eqref{YON} is the N system (\ref{N}), the invariant scaling transformation is
\[
\left(x,\,t,\,S,\,L\right)\to \left(\varepsilon\,x,\, \varepsilon^2\,t,\, \varepsilon^{-1}\,S,\,  \varepsilon^{-1}\,L\right)\,,
\]
where $\varepsilon\neq0$ is an arbitrary parameter.

%%%%%%%%%%%%%%%%%%%%%%%%%%%%%%%%%%%%%%%%%%%%%%%%%%%%%%%%%%%%%%%%%%%%
\section{Solutions of the YON model}
\label{sec:Sol}
%Consider the YON equation
%\begin{align}\label{system}
%\begin{aligned}
%&S_t -i S_{xx}-iS(i\alpha L_x + \alpha^2 L^2 - \beta L - 2 \alpha |S|^2) = 0,\\
%&L_t = 2(|S|^2)_x,
%\end{aligned}
%\end{align}
%where $\alpha$ and $\beta$ are arbitrary real parameters, $S:\Omega \rightarrow \mathbb{C}$, $L:\Omega\rightarrow \mathbb{R}$ and $\Omega\subset \mathbb{R}^2$ is an open set. In order to keep the system coupled, we will make the assumption that $\alpha^2 +\beta^2\neq 0$.
%\section{Traveling wave solutions}
Although system \eqref{YON} is integrable and hence it allows for solutions to be found by a variety of elegant and powerful solution techniques rooted into integrability theory, for the purpose of this paper we use instead an Ansatz to derive periodic and travelling wave solutions,
%without bringing up the heavy machinery of finite gaps theory and theta functions.
without resorting to heavier mathematical machineries.
%In order to obtain some explicit periodic and traveling wave solutions to \eqref{YON} using the Ansatz:
Our Ansatz naturally follows the form of the periodic and travelling wave solutions of the nonlinear Schr\"{o}dinger equation and it reads:
\begin{align}
\label{eq:ansatz}
    S(t,x) = s(z)e^{i(\phi(z)- \omega t)},\quad L(t,x) = \ell(z),
\end{align}
where $z=x-Vt$ for $V\in\mathbb{R}$, $V\neq0$,  $\omega\in\mathbb{R}$ and $s$, $\phi$, $\ell$ are real valued functions.
After substitution of the Ansatz into the second equation of \eqref{YON}, the resulting equation can be integrated in order to obtain $\ell$ in terms of $s$:
\begin{align}
\label{eq:F}
    %\boxed{F(z) = -\frac{2}{c}\phi^2(z) + K_1,}
    \ell(z) = -\frac{2}{V}s(z)^2 + c_1,
\end{align}
where $c_1$ is an arbitrary integration constant. We now make use of this new expression for $L$, substitute in the first equation of \eqref{YON} and separate it into real and imaginary parts in order to obtain the system

\begin{subequations}
\begin{align}
\label{eq:system1}
%\begin{aligned}
&s(z)\phi''(z)+2s'(z) \phi'(z) - \left[V+ \frac{4\alpha}{V}s(z)^2\right]s'(z)=0, \\
\label{eq:system2}
&s''(z) - s(z)\phi'(z)^2 + V\,s(z)\phi'(z) + (\alpha^2c_1^2-\beta c_1 + \omega)\,s(z) + \left(\frac{2\beta}{V} -\frac{4\alpha^2c_1}{V} - 2\alpha\right) s(z)^3 +\frac{4\alpha^2}{V^2}s(z)^5=0.
%\end{aligned}
\end{align}
\end{subequations}

for the functions $s$ and $\phi$.
%Multiplying equation (\ref{eq:system1}) by $\phi$, integrating with respect to $z$
Equation (\ref{eq:system1}) can be integrated with respect to $z$ after multiplication by $s$, yielding an expression for the first derivative of $\phi$ in terms of $s$:
%The first equation has a multiplier $\phi$, which means that we can integrate it after multiplication by $\phi$ and find the first integral
%\begin{align}
%    \phi^2(z)\psi'(z) - \frac{c}{2}\phi^2(z) - \frac{\alpha}{c}\phi^4(z) = K_2,
%\end{align}
%where $K_2$ is an integration constant. This means that we can obtain the first derivative of $\psi$ in terms of $\phi$:
\begin{align}
\label{eq:psiprime}
    %\boxed{\psi'(z) = \frac{\alpha}{c}\phi^2(z) + \frac{c}{2} + K_2\phi^{-2}(z).}
    \phi'(z) = \frac{\alpha}{V}\,s(z)^2 + \frac{V}{2} + c_2\,s(z)^{-2},
\end{align}
where $c_2$ is an integration constant.
We now plug this expression for $\phi'$ into equation \eqref{eq:system2}, obtaining

\begin{align}
\label{eq:psisecond}
    s''(z) +\frac{1}{4V}\left[V^3+4 V \left(\alpha ^2 c_1^2-\beta c_1+\omega \right)-8 \alpha c_2\right]\,s(z)+ \frac{2}{V}(\beta-\alpha V - 2\alpha^2c_1)\,s(z)^3 + \frac{3\alpha^2} {V^2}\,s(z)^5 - c_2^2\,s(z)^{-3}=0\,.
\end{align}

Equation (\ref{eq:psisecond}) can be integrated with respect to $z$ after multiplication by $s'$, resulting in a differential equation for $s'(z)^2$:

\begin{align}
\label{eq:psiprime2}
%(s'(z))^2 =- \frac{\alpha^2} {V^2}s(z)^6 +\frac{1}{V}(-\beta+\alpha V + 2\alpha^2c_1)s(z)^4-\frac{1}{4V}\left[V^3+4 V \left(\alpha ^2 c_1^2-\beta c_1+\omega \right)-8 \alpha c_2\right]s(z)^2+ 2c_3-c_2^2\, s(z)^{-2}\,,
s'(z)^2 = \displaystyle{-\frac{1}{4V}\left[V^3+4 V \left(\alpha ^2 c_1^2-\beta c_1+\omega \right)-8 \alpha c_2\right]s(z)^2}+\frac{1}{V}(-\beta+\alpha V + 2\alpha^2c_1)s(z)^4- \frac{\alpha^2} {V^2}s(z)^6 + 2c_3 -c_2^2\, s(z)^{-2},
\end{align}

where $c_3$ is an integration constant. Observe that for $\alpha=0$ the coefficient of $s(z)^6$ becomes zero and the equation simplifies leading to the Weierstra{\ss} elliptic function. The case $\alpha=0$ corresponds to the YO model, well studied in the literature, e.g. see \cite{CG2022,CCFMO2018} for some recent results about periodic and rational solutions, and literature therein.
\newline Here and thereafter we will assume $\alpha\neq 0$. Introducing the following change of variable
\begin{equation}
    \dfrac{V}{2\alpha} u(z) = s(z)^2-u_0\,,\quad u_0 = \dfrac{V(\alpha\,V + 2K_1 \alpha^2 - \beta)}{4\alpha^2}\,,\quad \alpha\neq 0
\end{equation}
%where
%\begin{equation}
%    u_0 = \dfrac{c(c\alpha + 2c_1 \alpha^2 - \beta)}{4\alpha^2}\,
%\end{equation}
equation \eqref{eq:psiprime2} becomes, without any loss of generality,
\begin{equation}
    \label{diffellip}
        u'(z)^2= -u(z)^4 + \mu_2 u(z)^2 + \mu_1 u(z) +\mu_0
\end{equation}
where

\begin{equation}
\begin{split}
\mu_0&=-\dfrac{V}{2\alpha^3} c^3 + \dfrac{4\alpha^2\omega - 8\alpha^2\,V^2 - \beta^2}{4\alpha^4} c^2 + \dfrac{\alpha^2\mu_1 + 2\alpha\,V^2(\mu_2+6\omega) - \alpha^2\,V^3 - 3\beta^2\,V}{2\alpha^3} c +\\
&+\frac14 \left\{ 2\mu_1\,V + V^2\left( 4\omega - \dfrac{\beta^2}{\alpha^2} \right) - \dfrac{\left[ \beta^2 - \alpha^2(\mu_2 + 4\omega) \right]^2}{\alpha^4} \right\}
 \end{split}
 \end{equation}
 
whereas $c$, $\mu_1$, $\mu_2$ are arbitrary constants. Note that the number of arbitrary constants is preserved.
The integration constants $c_1$, $c_2$  and $c_3$ can be rewritten in terms of $c$, $\mu_1$ and $\mu_2$ as follows
\begin{subequations}
\begin{equation}
    c_1 = \dfrac{c + \beta}{2\alpha^2},
\end{equation}
\begin{equation}
    c_2 = -\dfrac{V\left\{c^2 + 6c\alpha\,V + 2\beta^2 + \alpha^2\left[V^2-2(\mu_2+4\omega)\right] \right\}}{16\alpha^3}\,.
\end{equation}
\begin{equation}
    c_3 = \dfrac{V^3\left[ -2\alpha^3 \mu_1+ 2\alpha^2 \mu_2 (c+\alpha\,V)-(c+\alpha\,V)^3\right]}{32\alpha^4}\,.
% \left\{c^2 + 6Vc\alpha + 2\beta^2 + \alpha^2\left[V^2-2(\mu_2+4\omega)\right] \right\}}{16\alpha^3}\,.
\end{equation}
\end{subequations}
As for $\phi(z)$, from \eqref{eq:psiprime} we obtain the quadrature
\begin{equation}
\phi'(z)= \dfrac{\alpha\left[ \alpha\,V^2 - c\,V + \alpha(\mu_2+4\omega) \right] - \beta^2 + 2\alpha u(z) [c + 2\alpha\,V + \alpha u(z)]}{2\alpha [c + \alpha\,V + 2\alpha u(z)]}
\end{equation}
For $\mu_1=0$ equation \eqref{diffellip} admits periodic solutions in terms of Jacobi elliptic functions, which we are going to discuss below.
\subsection{Jacobi elliptic sine solution}
Let us assume that $u(z)$ has the form
\begin{equation}
\label{eq:Jes}
    u(z) = \gamma_0 + \gamma_1 \operatorname{sn}\Big(a(z-z_0),m\Big)\,,
\end{equation}
where $\operatorname{sn}(z)$ denotes the Jacobi elliptic sine of $z$, where $\gamma_0$, $\gamma_1$, $a$, $z_0$ and $m$ are real parameters, and $0\leq m\leq 1$.
Inserting \eqref{eq:Jes} into \eqref{diffellip} with $\mu_1=0$ and playing with the properties of the Jacobi elliptic functions, one obtains a polynomial of degree four in $\operatorname{sn}(a(z-z_0),m)$ equated to zero. Setting the coefficients of each power of $\operatorname{sn}(z)$ to zero, one obtains a set of algebraic equations for the parameters $\omega$, $m$, $a$, $\gamma_0$, $\gamma_1$, $\mu_2$ and $c$.  In particular, a relation for $\omega$ in terms of the other parameters can be found by setting the constant term of the polynomial to zero.  This latter relation returns a non-real $\omega$ for any choice of the other parameters, and therefore this excludes the existence of a $\operatorname{sn}$-solution of the form \eqref{eq:Jes} starting from Ansatz \eqref{eq:ansatz}.
%%%%%%%%%%%%%%%%%%%%%%%%%%%%%%%%%%%%%%%%%%%%%%%%%%%%%%%%%%%%%%%%
\subsection{Jacobi elliptic cosine solution}
Proceeding as above but with the Jacobi elliptic cosine $\operatorname{cn}(z)$ replacing $\operatorname{sn}(z)$ in \eqref{eq:Jes}, we obtain the following solution to \eqref{diffellip} in:
\begin{equation}
\label{eq:Jec}
    u(z) =  ma \operatorname{cn}\Big(a(z-z_0),m\Big)\,,
\end{equation}
with
\begin{subequations}
\begin{equation}
    \mu_2=(2m^2-1)a^2\,,\quad c=b-\alpha\,V\,,
\end{equation}
and

\begin{equation}
    \label{omegacn}
    \omega = \dfrac{2\beta^2 + 2\alpha^2\left[a^2(1-2m^2) - 2V^2\right] + 4\alpha\,V b +  b^2 \pm
    \sqrt{(b - 2 a m \alpha)\,(b + 2 a m \alpha)\,\left[b^2 + 4 a^2 (1 - m^2) \alpha^2\right]} }{8\alpha^2}\,,
    %\sqrt{16m^2(m^2-1)\alpha^4a^4 + 4(1-2m^2)\alpha^2a^2 b^2+ b^4}}{8\alpha^2}\,,
\end{equation}

\end{subequations}
where $m$, $z_0$, $a\neq 0$ and $b$ are real parameters, with $0\leq m\leq 1$.
Replacing the expression of $u(z)$ into the original Ansatz \eqref{eq:ansatz}, we obtain the following solution of the YON system:
\begin{subequations}
\label{eq:Jecsol}
\begin{align}
    S(x,t)&= \frac12 e^{i(\phi(z-\omega t)} \sqrt{\dfrac{V\left[ b + 2m\alpha a \operatorname{cn}\Big( a(z-z_0),m\Big)\right]}{\alpha^2}}\,,\\
    L(x,t)&= \dfrac{\beta - \alpha\,V - 2m\alpha a \operatorname{cn}\Big(a(z-z_0),m\Big)}{2\alpha^2}\,,\quad z=x-Vt\,,
\end{align}
where $\phi(z)$ satisfies the quadrature

\begin{equation}
\label{eq:Jecphi}
    \phi'(z)=\dfrac{1}{4\alpha} \left\{ b + 2\alpha\left[ V+m a \operatorname{cn}\Big(a(z-z_0),m\Big) \right]\pm
    \dfrac{
    %\sqrt{16m^2(m^2-1)\alpha^4 a^4 + 4(1-2m^2)\alpha^2 a^2 b^2+ b^4}
    \sqrt{(b - 2 a m \alpha)\,(b + 2 a m \alpha)\,\left[b^2 + 4 a^2 (1 - m^2) \alpha^2\right]}
    }
    {2m\alpha a\operatorname{cn}\Big(a(z-z_0),m\Big) +  b} \right\}\,,
\end{equation}

\end{subequations}
where the sign in front of the square root is the same sign chosen for $\omega$. Observe that, in addition to the coupling parameters $\alpha$ and $\beta$ in the YON model \eqref{YON}, the solutions \eqref{eq:Jecsol} features five real parameters, namely $a$, $b$, $m$, $V$ and $z_0$, with a sixth real, arbitrary parameter coming from the integration of \eqref{eq:Jecphi}.

For this to work, we need that all $s$, $\phi$, $\omega$ be real. In order to assure that, we need to check the sign inside all the square roots involved, which gives us constraints on the parameters, namely
\begin{equation}
    \label{constraints_CN}
    \left| \dfrac{ b}{\alpha a} \right| \ge 2m\,,\quad V b \ge 0\,.
\end{equation}
Furthermore, in the special case $m=1$, the value $ b=0$ is allowed as long as $\alpha\,V a>0$, see Section \ref{sec:solitons}.

The short wave $|S|$ oscillates between the values
$$
\dfrac12 \sqrt{\dfrac{V(b+2m\alpha a)}{\alpha^2}}
\qquad\mbox{and}\qquad
\dfrac12 \sqrt{\dfrac{V(b-2m\alpha a)}{\alpha^2}}\,,
$$
that is to say, $|S|^2$ oscillates with amplitude $\left|Vma/\alpha\right|$,
while the long wave $L$ oscillates between the values
$$
\dfrac{\beta-\alpha\,V-2m\alpha a}{2\alpha^2}
\qquad\mbox{and}\qquad
\dfrac{\beta-\alpha\,V+2m\alpha a}{2\alpha^2}\,,
$$
that is, with amplitude $2\left| m a/\alpha\right|$.
The cn-solution is periodic in $x$ with period $\left|4 K(m)/a\right|$ and in $t$ with period $\left|4 K(m)/(aV)\right|$, where $K(m)$ is the complete elliptic integral of the first kind of $m$,
\begin{equation}
    K(m) = \int_0^{\frac{\pi}{2}} \dfrac{d\theta}{\sqrt{1-m\sin^2\theta}} = \int_0^1 \dfrac{dt}{\sqrt{(1-t^2)(1-mt^2)}}\,.
\end{equation}
Moreover, we observe that there are two special solutions corresponding to the particular choices $m=0$ and $m=1$. If $m=0$, the elliptic cosine reduces to the trigonometric cosine and the solution becomes a plane wave. If $m=1$ the elliptic cosine reduces to the hyperbolic secant, leading to a localised solution, treated in Section \ref{sec:solitons}.
\begin{figure}[h!]
    \centering
\begin{subfigure}{0.4\textwidth}
    \centering
    \includegraphics[width=\textwidth]{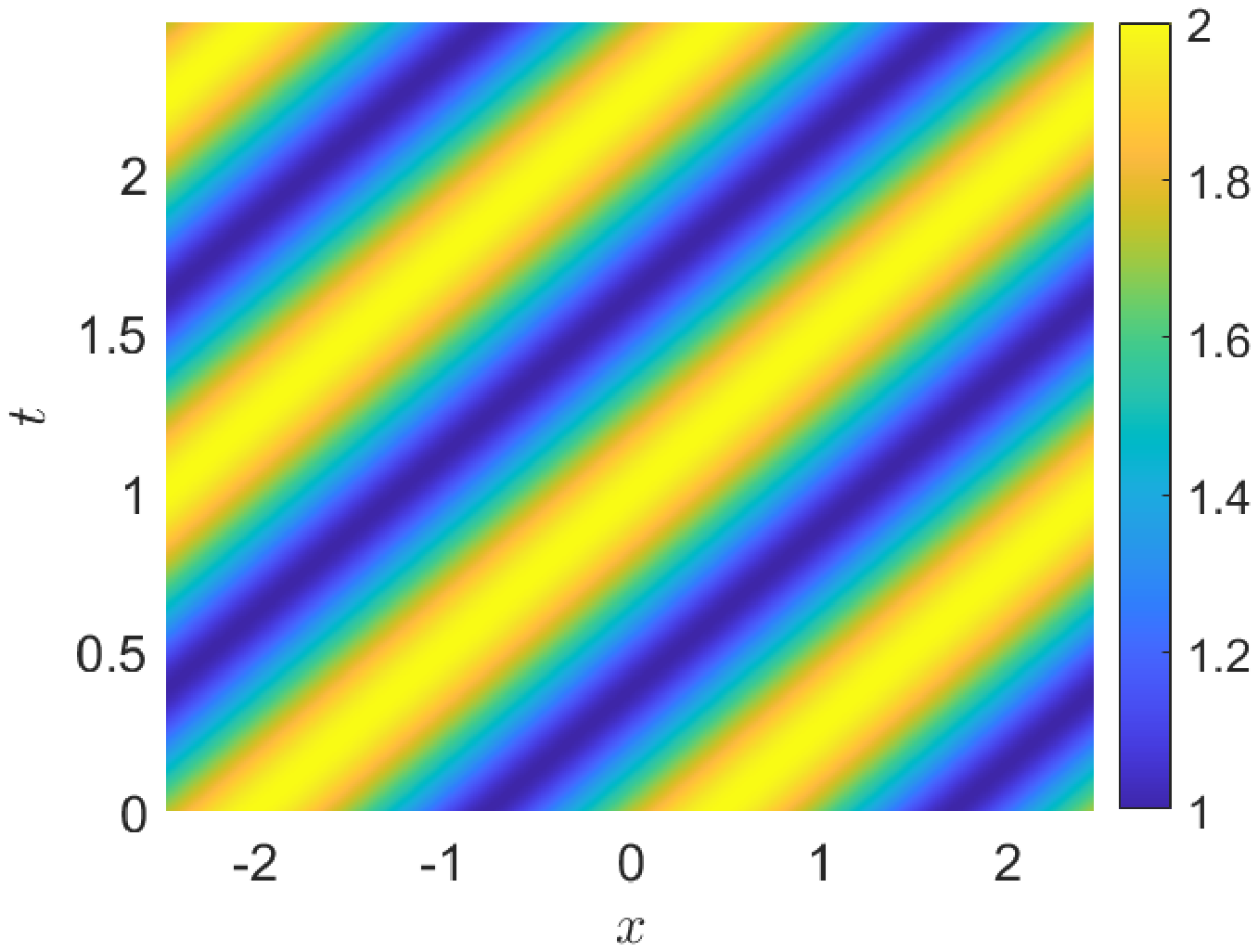}
    \subcaption{Short wave $|S|$.}
\end{subfigure}
\hfill
\begin{subfigure}{0.4\textwidth}
\centering
    \includegraphics[width=\textwidth]{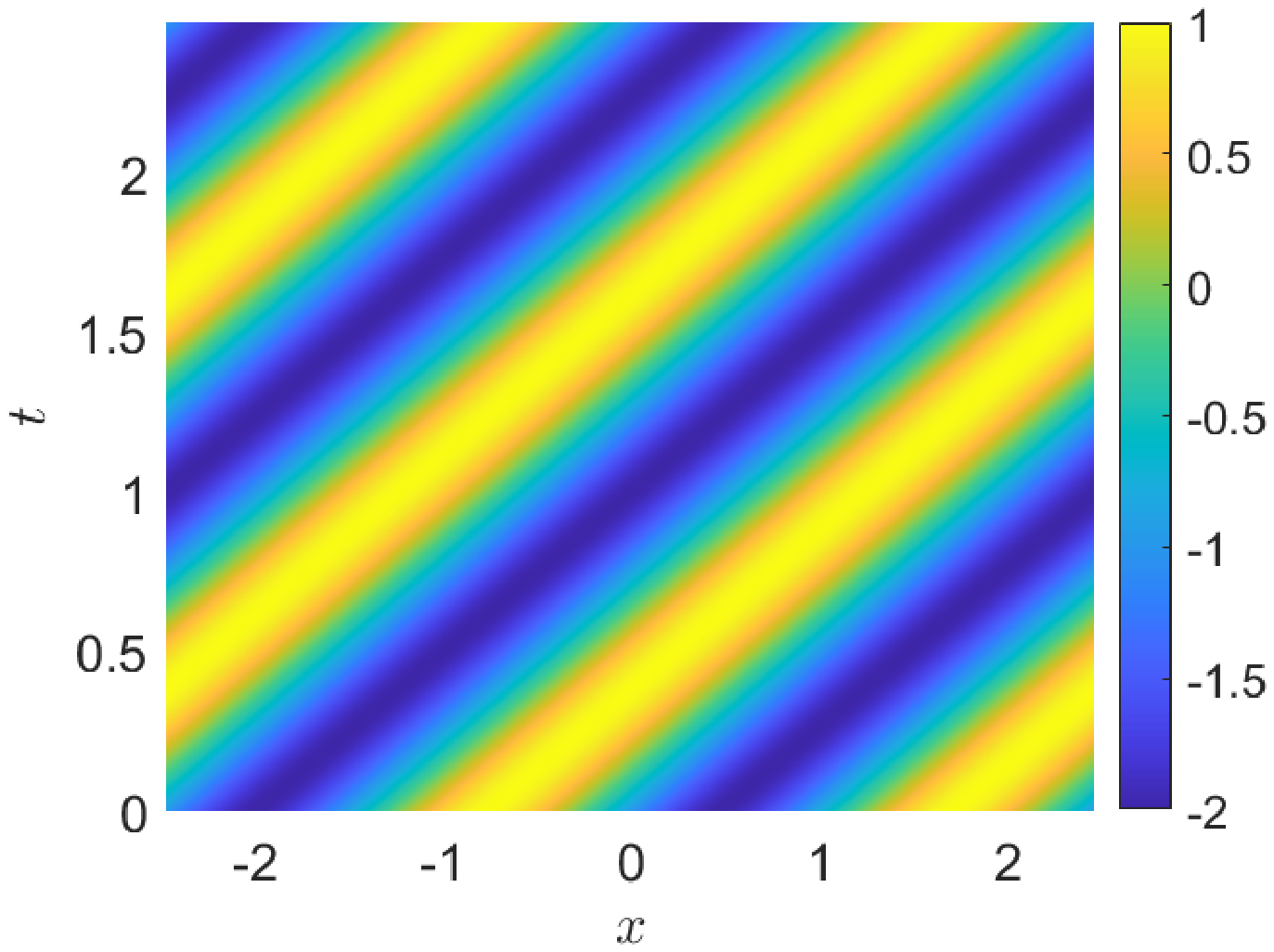}
    \subcaption{Long wave $L$.}
\end{subfigure}
    \caption{Elliptic cosine solution with $\alpha=1$, $\beta=2$, $V=1.2$, $ b=5$, $ a=1.3$, $z_0=0$, $m=0.5$.}
\end{figure}
%%%%%%%%%%%%%%%%%%%%%%%%%%%%%%%%%%%%%%%%%%%%%%%%%%%%%%%%%%%%%%%%
\subsection{Jacobi delta amplitude solution}
%Finally, let us assume that $U(z)$ has the form
%\begin{equation}
%    U(z) = \eta_0 + \eta_1 \operatorname{dn}(\delta(z-z_0),m)\,,
%\end{equation}
%where $\operatorname{dn}(z)$ denotes the delta amplitude of $z$.
Proceeding as above with the Jacobi delta amplitude  $\operatorname{dn}(z)$ replacing $\operatorname{sn}(z)$ in \eqref{eq:Jes}, we obtain the following solution to \eqref{diffellip}
\begin{equation}
\label{eq:Jed}
    u(z) =  \frac{ a}{m} \operatorname{dn}\Big( a(z-z_0),m\Big)\,,
\end{equation}
with
\begin{subequations}
\begin{equation}
    \mu_2=\left(\frac{2}{m^2} - 1\right) a^2\,,\quad c= b-\alpha\,V\,,
\end{equation}
and

\begin{equation}
       \omega= \dfrac{1}{8m^2\alpha^2} \Bigg\{ 2\alpha^2\big[-2V^2m^2+(m^2-2) a^2\big] + 4V m^2 \alpha  b + m^2(2\beta^2 +  b^2)
        \pm
        \sqrt{(m b - 2 a \alpha)\,(m b + 2 a \alpha) \left[m^2 b^2 - 4 a^2 (1 - m^2) \alpha^2\right]}
        %\sqrt{16(1-m^2)\alpha^4 a^4 + 4m^2(m^2-2)\alpha^2 a^2 b^2 + m^4 b^4}
        \Bigg\}\,,
\end{equation}

\end{subequations}
where $m$, $z_0$, $ a\neq 0$ and $b$,  are real parameters, with $0\leq m\leq 1$.
%where we have set $\eta_0=0$, $\eta_1=\frac{\delta}{m}$, $\mu_1=0$, $\mu_2=\left(\frac{2}{m^2} - 1\right)\delta^2$ and $K_0=\kappa_0-c\alpha$.

The solution in this case has the form
\begin{subequations}
\label{eq:Jedsol}
\begin{align}
&    S(x,t)= \frac12 e^{i(\phi(z)-\omega t)} \sqrt{\dfrac{V\left[m b + 2\alpha a \operatorname{dn}\Big( a(z-z_0),m\Big)\right]}{m\alpha^2}}\,,\\
&    L(x,t)= \dfrac{m(\beta -\alpha\,V)-2\alpha a \operatorname{dn}\Big( a(z-z_0),m\Big)}{2m\alpha^2}\,,\quad z=x-Vt
\end{align}
where $\phi$ satisfies the quadrature

\begin{equation}
\label{eq:Jedphi}
\phi'(z) =
\dfrac{1}{4m\alpha}\left[2Vm\alpha + m b + 2\alpha a \operatorname{dn}\Big( a(z-z_0),m\Big) \pm
%\frac{\sqrt{16(1-m^2)\alpha^4 a^4 + 4m^2(m^2-2)\alpha^2 a^2 b^2 + m^4 b^4}
\frac{\sqrt{(m b - 2 a \alpha)\,(m b + 2 a \alpha) \left[m^2 b^2 - 4 a^2 (1 - m^2) \alpha^2\right]}
}{2\alpha a\operatorname{dn}\Big( a(z-z_0),m\Big) +m b}\right]\,,
\end{equation}

\end{subequations}
where the sign in front of the square root is the same sign chosen for $\omega$. Similarly to the Jacobi elliptic solutions \eqref{eq:Jecsol}, observe that the solutions \eqref{eq:Jedsol} features five real parameters, namely $a$, $b$, $m$, $V$ and $z_0$, with a sixth real, arbitrary parameter coming from the integration of \eqref{eq:Jedphi}.

Again, checking the square roots that appear for having real solutions, we get the following constraints on the parameters:
\begin{equation}
    2\sqrt{\dfrac{1-m}{m^2}}\le \dfrac{ b}{\alpha a} \le 2\sqrt{\dfrac{1-m^2}{m^2}}\,,\quad \alpha\,V a>0\,,\quad 0<m\le1\,.
\end{equation}
It also allows the special values $\frac{ b}{\alpha a}=-2\sqrt{\frac{1-m}{m^2}}$ and $\frac{ b}{\alpha a}=\frac{2}{m}$, for $0<m\le 1$. We discuss the special case $m=1$ in Section \ref{sec:solitons}.
% {\color{green} These are the conditions for the actual square roots that appear in the general solution of $\phi$ and $\omega$, I guess different solutions may be possible when $m=0$ and $m=1$, but then one needs to rewrite $\omega$ and the quadrature for $\phi$.}

The dn-solution has periodicity for $L$ and $|S|$ with period $\left|2 K(m)/a\right|$, where $K(m)$ is the complete elliptic integral of the first kind of $m$, while the phase of $S$ has a period $\left|4 K(m)/a\right|$.
%{\color{red} Do we need to specify $ a=0$ in this case?} {\color{green} No, dn(z,0)=1 for all z, but $m=0$ is not allowed by the form of the solutions we use.}
%For the case $m=1$, as for the elliptic cosine the period diverges and the solution becomes localised, reducing to a bright soliton if $ b=0$ and $V=\frac{\beta}{\alpha}$ and to a dark soliton otherwise. In the mixed case $ b=0$, $V\neq\frac{\beta}{\alpha}$ the solution is bright in $S$ and dark in $L$, while with $ b\neq0$, $V=\frac{\beta}{\alpha}$ it is dark in $S$ and bright in $L$.

The short wave $|S|$ oscillates between the values
$$
\dfrac12 \sqrt{\dfrac{V[2\alpha a (1-m)^{\nicefrac12} + mb]}{m\alpha^2}}
\qquad\mbox{and}\qquad
\dfrac12 \sqrt{\dfrac{V(2\alpha a + mb)}{m\alpha^2}}\,,
$$
that is to say, the oscillations in $|S|^2$ have an amplitude $\left|V a (1-\sqrt{1-m})/(2\alpha)\right|$,
while the long wave $L$ oscillates between the values
$$
\dfrac{m(\beta -\alpha\,V)-2\alpha a}{2m\alpha^2}
\qquad\mbox{and}\qquad
\dfrac{m(\beta -\alpha\,V)-2\alpha a \sqrt{1-m}}{2m\alpha^2}\,,
$$
that is, with an amplitude $\left| a(1-\sqrt{1-m})/\alpha \right|$.
\begin{figure}[h!]
    \centering
    \begin{subfigure}{0.4\textwidth}
    \centering
    \includegraphics[width=\textwidth]{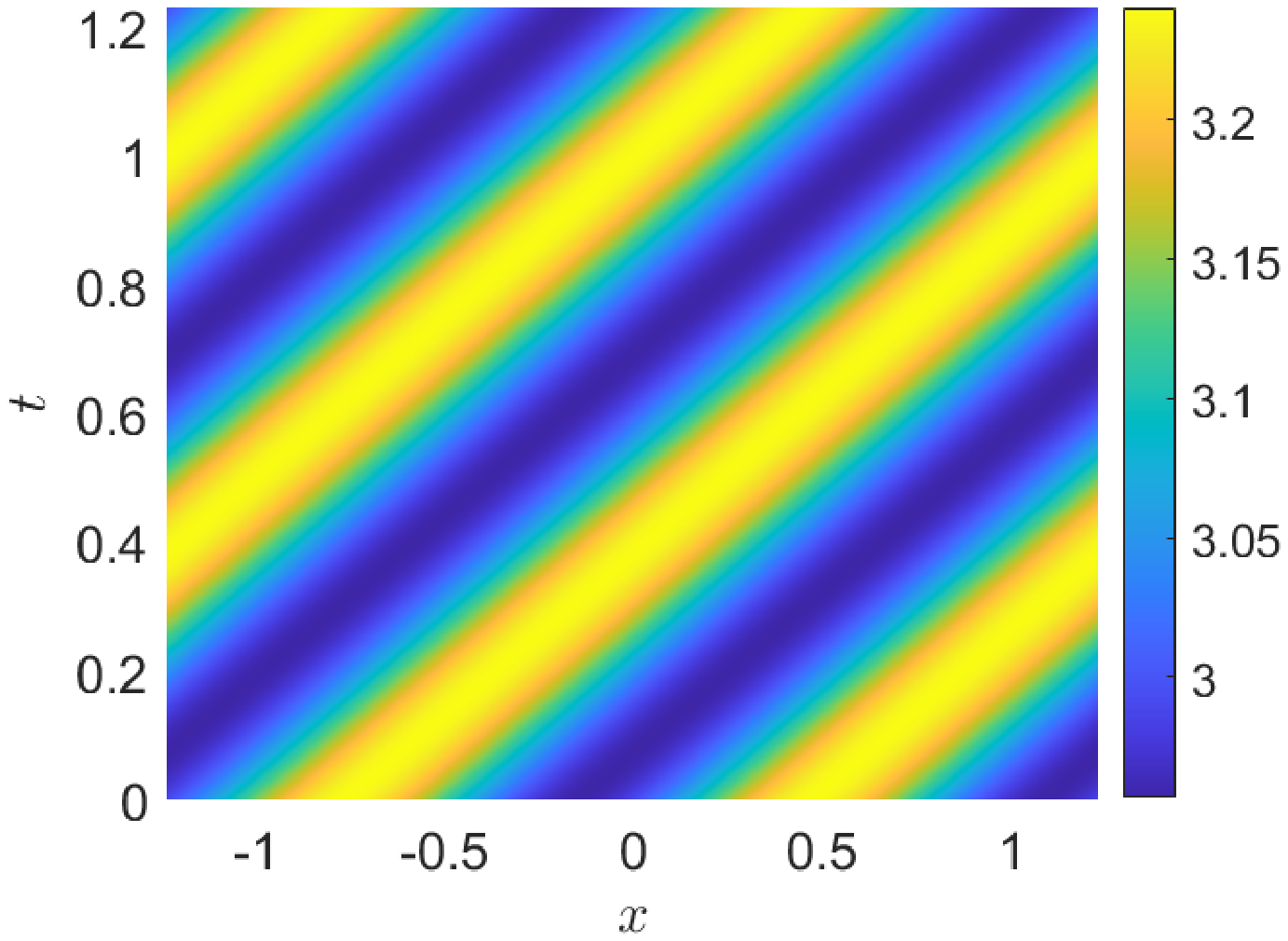}
    \subcaption{Short wave $|S|$.}
\end{subfigure}
\hfill
\begin{subfigure}{0.4\textwidth}
\centering
    \includegraphics[width=\textwidth]{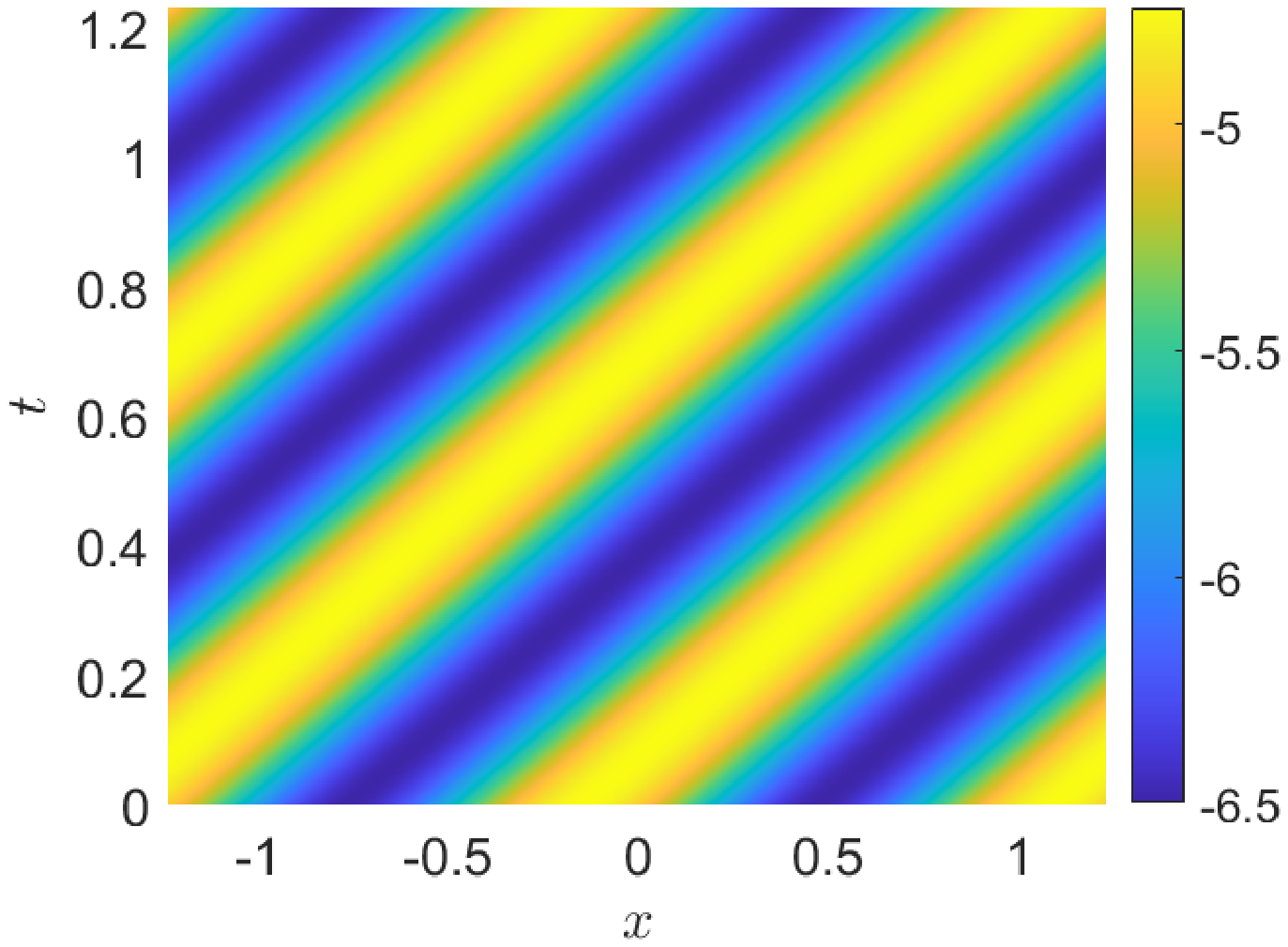}
    \subcaption{Long wave $L$.}
\end{subfigure}
    \caption{Delta amplitude solution with $\alpha=1$, $\beta=1$, $V=2$, $ b=9$, $ a=3$, $z_0=-2$, $m=0.5$.}
    \label{fig:ellipticDN}
\end{figure}
%%%%%%%%%%%%%%%%%%%%%%%%%%%%%%%%%%%%%%%%%%%%%%%%%%%%%%%%%%%%%%%
\subsection{Travelling waves: solitons}
\label{sec:solitons}
%
%{\color{red} General solution and bright solution}
The choice $m = 1$ in \eqref{eq:Jec} makes the period of the elliptic cosine diverge, so the solution becomes localised. The corresponding solutions are solitons, both of dark and bright type.

The solution for $m=1$, which for a generic choice of parameters corresponds to a dark soliton, has the form
\begin{subequations}

\begin{align}
&    S(x,t)= \frac12 e^{i(\phi(z)-\omega t)} \sqrt{\dfrac{V\left[ b + 2\alpha a \operatorname{sech}\Big( a(z-z_0)\Big)\right]}{\alpha^2}}\,,\\
&    L(x,t)= \dfrac{\beta - \alpha\,V - 2\alpha a \operatorname{sech}\Big( a(z-z_0)\Big)}{2\alpha^2}\,,\quad z=x-Vt\,,\\
&    \omega= \dfrac{1}{8\alpha^2} \Big[2\beta^2 - 2\alpha^2(2V^2+ a^2) + 4V \alpha  b +  b^2
        \pm \sqrt{ b^4-4\alpha^2 a^2 b^2 }\Big]\,,
\end{align}

with

\begin{align}
\phi(z) = &\dfrac{z-z_0}{4\alpha} \left( 2\alpha\,V +  b \pm\mathrm{sgn}(b)\,\sqrt{ b^2 - 4\alpha^2 a^2} \right) + \arctan \Big(\tanh\Big(\frac{ a}{2}(z-z_0)\Big) \Big) \notag \\
    &\mp\mathrm{sgn}(b)\,\arctan \Bigg(\dfrac{(b-2\alpha a) \tanh\Big( \frac{ a}{2}(z-z_0)\Big)}{\sqrt{ b^2-4\alpha^2 a^2}}\Bigg) + \phi_0\,,
\end{align}

\end{subequations}
where $\phi_0$ is an arbitrary phase, and the sign function $\mathrm{sgn}$ satisfies $\mathrm{sgn}(0)=0$.
%{\color{red} Formula (28d) needs checking. Condition on the denominator should also be added.}
As a consequence of \eqref{constraints_CN}, and as it can be observed from the formulae above, the general condition on the parameters for the validity of the soliton solution, when $b\neq0$, is
\begin{equation}
    \left|\dfrac{b}{\alpha a}\right| \ge 2\,.
\end{equation}
The special choice $b=0$ (see below) is also allowed by the system, though the resulting solution has the phase:
$$
\phi(z)= \dfrac{V\,(z-z_{0})}{2} + \arctan \Big(\tanh\Big(\frac{ a}{2}(z-z_0)\Big)\Big) + \phi_0\,.
$$
The square of the short wave, $|S|^2$, has an amplitude $\dfrac12 \left| \dfrac{V a}{\alpha} \right|$ over the background $\left| \dfrac{Vb}{4\alpha} \right|$, while the long wave $L$ has an amplitude $-\dfrac{a}{\alpha}$ over the background $\dfrac{\beta-\alpha\,V}{2\alpha^2}$. Note that both amplitudes and the background of $S$ do not depend on $\beta$ at all, while they all depend inversely on $\alpha$.

By construction, both $S(x,0)$ and $L(x,0)$ are centred at $x=z_0$, while $S(x,t_0)$ and $L(x,t_0)$ for a given $t_0$ are both centred at $x=z_0+V t_0$.

Whenever $b=0$, $S$ has zero background, and whenever $V=\beta/\alpha$, $L$ has zero background too. Both equalities being true means having a bright soliton solution, both being false leads to a dark soliton solution, while the cases where one is true and the other is not lead to mixed bright-dark solutions.

The resulting formula for the bright case is
\begin{subequations}
\label{bright}
\begin{align}
&    S(x,t)= \frac{\sqrt{2}}{2} e^{i(\phi(z)-\omega t)} \sqrt{\dfrac{\beta a \operatorname{sech}\Big(a(z-z_0)\Big)}{\alpha^2}}\,,\\
&    L(x,t)= -\dfrac{ a \operatorname{sech}\Big(a(z-z_0)\Big)}{\alpha}\,,\quad z=x-\dfrac{\beta}{\alpha}t\,,\\
&    \omega= -\dfrac{\beta^2+\alpha^2 a^2}{4\alpha^2}\,,\\
&    \phi(z)= \dfrac{\beta (z-z_{0})}{2\alpha} + \arctan \Big(\tanh\Big(\frac{ a}{2}(z-z_0)\Big)\Big) + \phi_0\,.
\end{align}
\end{subequations}
The same procedure can be carried out by taking $m=1$ in the dnoidal solution \eqref{eq:Jed}. The solitons obtained in this way have the exact same formula as the ones obtained from the cnoidal case.

\begin{figure}[h!]
    \centering
    \begin{subfigure}{0.4\textwidth}
    \centering
    \includegraphics[width=\textwidth]{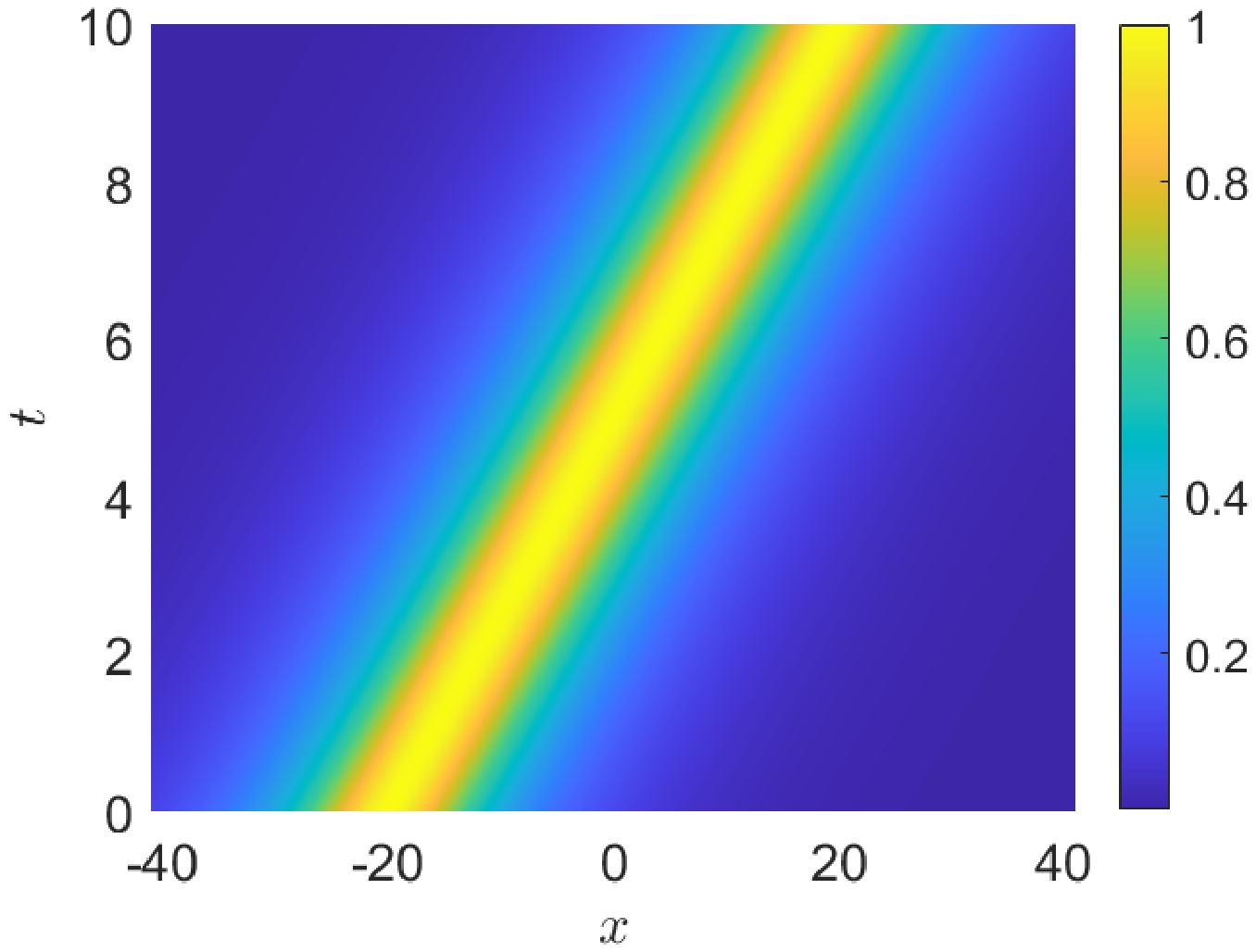}
    \subcaption{Short wave $|S|$.}
\end{subfigure}
\hfill
\begin{subfigure}{0.4\textwidth}
\centering
    \includegraphics[width=\textwidth]{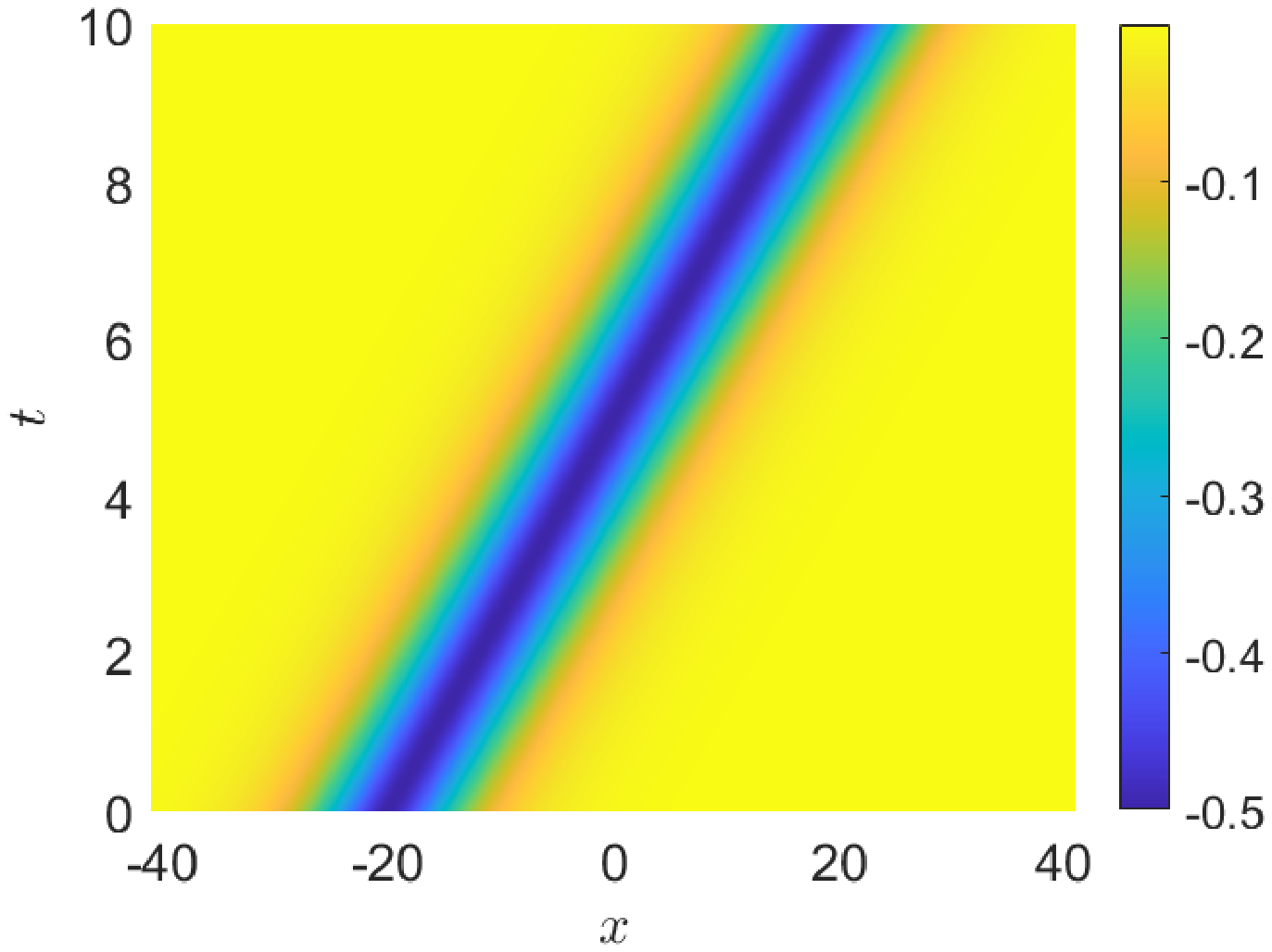}
    \subcaption{Long wave $L$.}
\end{subfigure}
    \caption{Bright soliton solution, with $\alpha=0.5$, $\beta=2$, $a=0.25$, $b=0$, $V=4$, $z_0=-20$.}
    \label{fig:soliton}
\end{figure}

%%%%%%%%%%%%%%%%%%%%%%%%%%%%%%%%%%%%%%%%%%%%%%%%%%%%%%%%%%%%%%%
\subsection{Travelling waves: rational solution}
\label{sec:rational}
%Consider \eqref{eq:psiprime2} and write it in a more convenient way for us to find a rational solution. After multiplication by $s'$, we obtain the following quadrature:
%\begin{align}\label{quadrature}
%	\begin{aligned}
%    (s'(z))^2 =& \displaystyle{\frac{-\frac{1}{4V}\left[V^3+4 V \left(\alpha ^2 c_1^2-\beta c_1+\omega \right)-8 \alpha c_2\right]s(z)^4}{s(z)^2}}\\
%    &\displaystyle{+\frac{\frac{1}{V}(-\beta+\alpha V + 2\alpha^2c_1)s(z)^6- \frac{\alpha^2} {V^2}s(z)^8 + 2c_3s(z)^2 -c_2^2}{s(z)^2}},
%    \end{aligned}
%\end{align}
%where $c_3$ is another integration constant.
By letting $c_1=c_2=c_3=0$ in \eqref{eq:psiprime2} it reads
\begin{equation}
\label{eq:sprime2_2}
s'(z)^2 = - \frac{\alpha^2}{V^2}s(z)^2 \left[s(z)^4-\frac{V}{\alpha^2}(\alpha V - \beta) + \frac{V^2}{3\alpha^4}(\alpha V -\beta)^2\right],
\end{equation}
where $\omega$ is given below. An integration process now yields
\begin{align*}
    \int_{s}^{0} \frac{3\sqrt{3}\,\alpha^2|\alpha|}{(V(\alpha V-\beta)-3\alpha^2\zeta^2)^{3/2}}d\zeta= \pm\sqrt{\frac{\alpha^2}{V^2}}z\,.
\end{align*}
%and then the function $s(z)$ is given by
%\begin{align}
%    \begin{aligned}
%    s(z) &= \frac{\sqrt{3}V|\alpha|}{3\alpha^2|V|}(\alpha V-\beta)z\sqrt{\frac{V(\alpha V-\beta)}{9\alpha^2 + (\alpha V-\beta)^2z^2}}.
%    \end{aligned}
%\end{align}
%Substitution of $s(z)$ into the expressions for $\ell(z)$ and $\phi(z)$ leads to the solution
Computing the integral, solving with respect to $s(z)$ and substituting into the expressions for $\ell(z)$ and $\phi(z)$ leads to the solution
\begin{subequations} %\frac{\sqrt{3}V|\alpha|}{3\alpha^2|V|}
\begin{align}\label{eq:rationalsol}
&    S(x,t) = \frac{z}{\sqrt{3}\,\alpha}\,e^{i(\phi(z)-\omega t)}\sqrt{\frac{V\,(\alpha V-\beta)^{3}}{9\alpha^2 + (\alpha V-\beta)^2z^2}},\\
&    L(x,t)= -\frac{2(\alpha V-\beta)^3 z^2}{3\alpha^2[9\alpha^2 + (\alpha V-\beta)^2z^2]},\quad z=x-Vt,\\
&    \omega = \frac{\alpha^2V^2-8\alpha \beta V+4\beta^2}{12 \alpha^2}\,,\\
&    \phi(z) = \arctan\left(\frac{3\alpha}{(\alpha V -\beta)z}\right) + \left(\frac{V}{2} +\frac{\alpha V -\beta}{3\alpha}\right)z + \phi_0,
\end{align}
\end{subequations}
where $\phi_0$ is an arbitrary phase, and the constraint $V(\alpha V-\beta)>0$ is assumed.

The short wave $|S|$ is a dark rational solution with an amplitude depression of $\sqrt{V(\alpha V-\beta)/(3\alpha^2)}$ propagating on the non-vanishing background $\sqrt{V(\alpha V-\beta)/(3\alpha^2)}$, whereas the long wave $L$ has an amplitude $2(\alpha V-\beta)/(3\alpha^2)$ over the asymptotic background $-2(\alpha V-\beta)/(3\alpha^2)$.

\begin{figure}[h!]
    \centering
    \begin{subfigure}{0.4\textwidth}
    \centering
    \includegraphics[width=\textwidth]{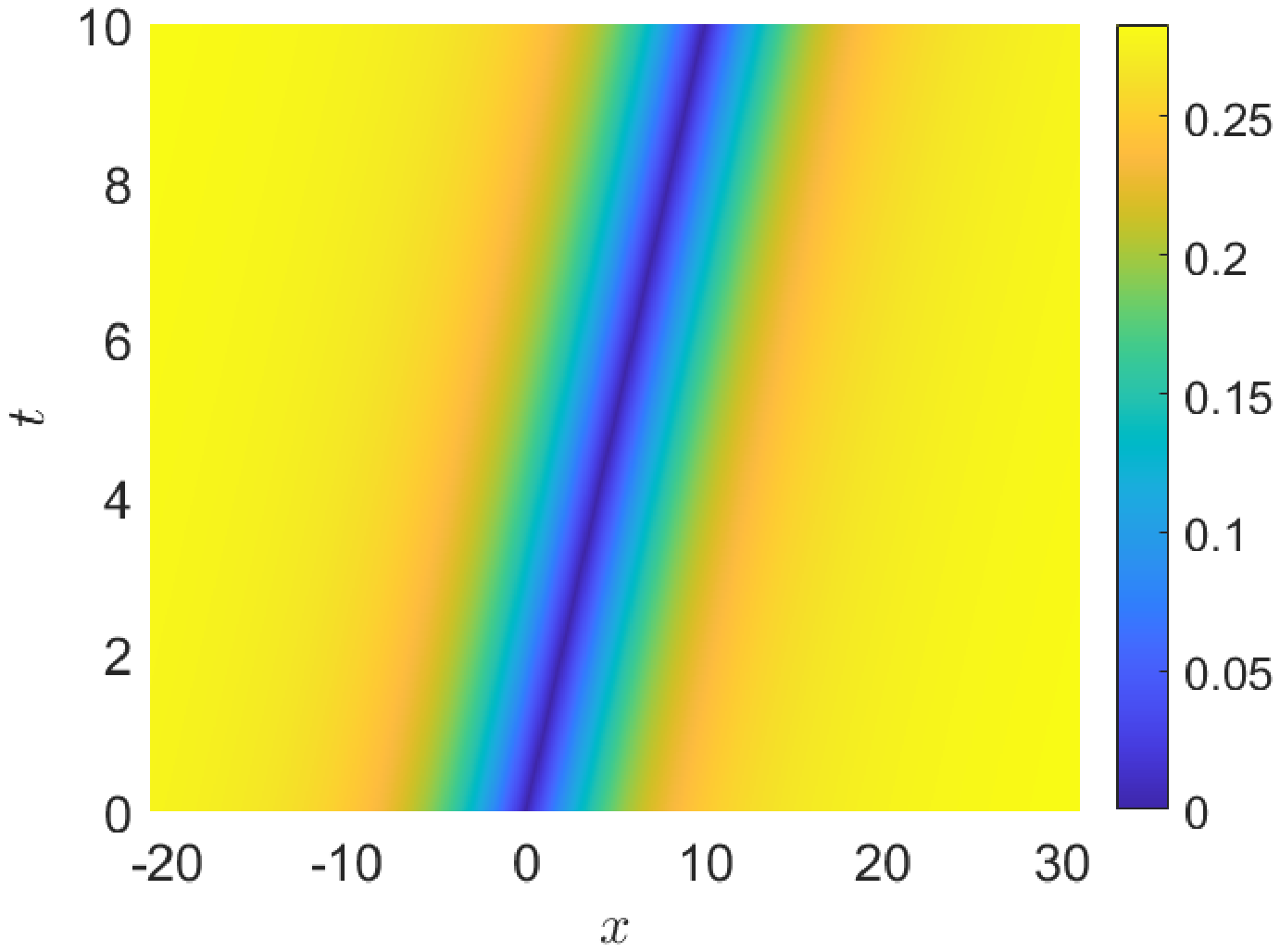}
    \subcaption{Short wave $|S|$.}
\end{subfigure}
\hfill
\begin{subfigure}{0.4\textwidth}
\centering
    \includegraphics[width=\textwidth]{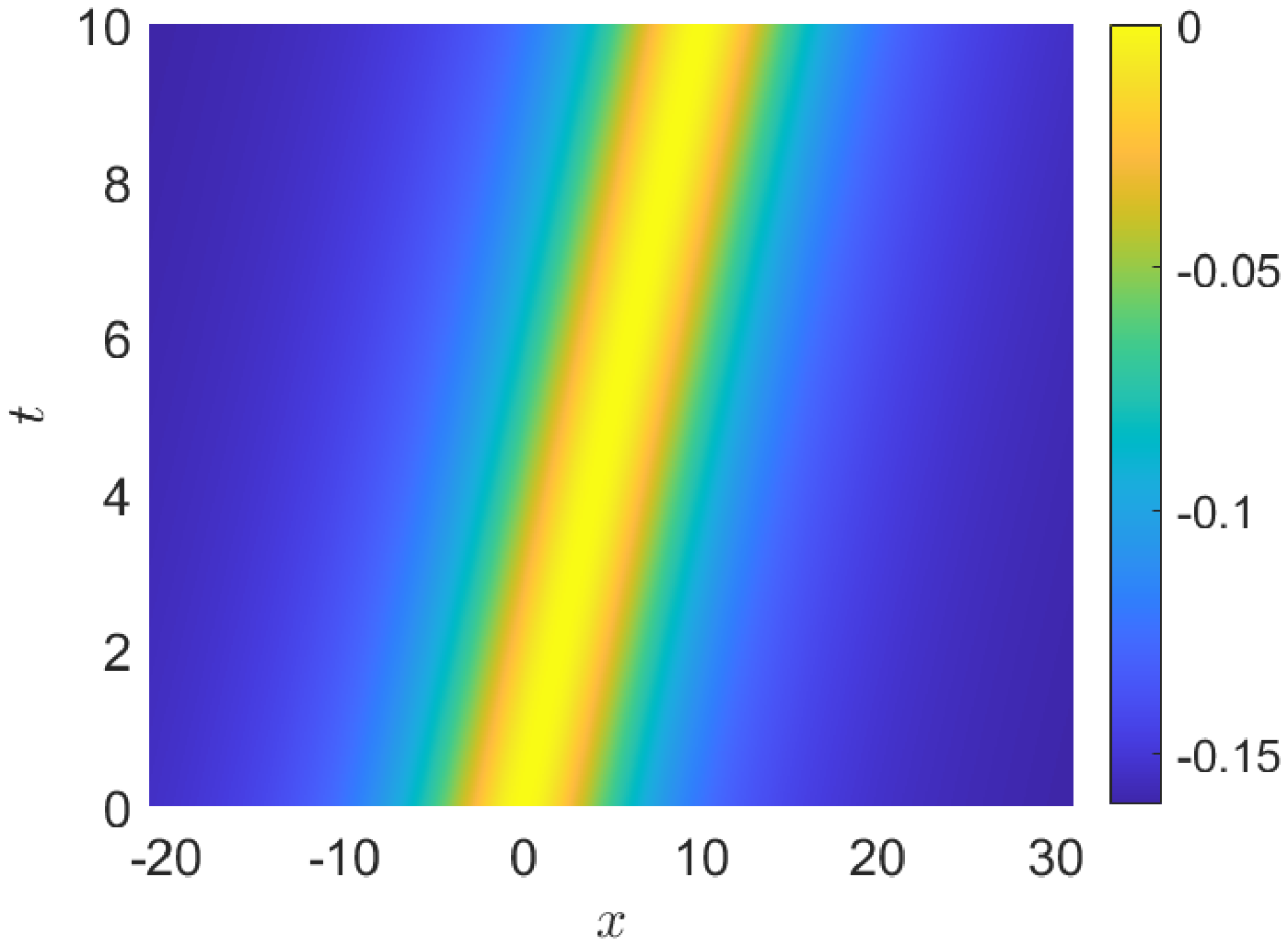}
    \subcaption{Long wave $L$.}
\end{subfigure}
    \caption{Rational solution with $\alpha=2$, $\beta=1$, $V=1$.}
    \label{fig:soliton}
\end{figure}
In the case where $\beta=0$ and we are back to the Newell system \eqref{N}, the solution is simplified to
\begin{subequations}
\begin{align}\label{eq:rationalsolN}
&    S(x,t) = \frac{z}{\sqrt{3}\,\alpha}\,e^{i(\phi(z)-\omega t)}\sqrt{\frac{\alpha\,V^{4}}{9 + V^2z^2}},\\
&    L(x,t)= -\frac{2V^3z^2}{3\alpha[9 + V^2z^2]},\quad z=x-Vt,\\
&    \omega = \frac{V^2}{12}\,,\\
&    \phi(z) = \arctan\left(\frac{3}{Vz}\right) + \left(\frac{V}{2} +\frac{\alpha V}{3\alpha}\right)z + \phi_0,
\end{align}
\end{subequations} where $\alpha>0$. To the best of our knowledge, this is the first time such a solution is derived for \eqref{N}.

Another solution can be obtained for $V(\alpha V-\beta)>0$ but now for $\omega=-c^2/4$, still with $c_1=c_2=c_3=0$. The quadrature  \eqref{eq:psiprime2} is now rewritten as
$$s'(z)^2 = -\frac{\alpha^2}{V^2} s(z)^4\left[s(z)^2 -\frac{V}{\alpha^2}(\alpha V-\beta)\right],$$
and integration leads to the solutions
\begin{subequations}
\begin{align}
\label{eq15}
&    S(x,t) = e^{i(\phi(z)-\omega t)}\sqrt{\frac{V(\alpha V-\beta)}{\alpha^2+(\alpha V -\beta)^2z^2}},\\
&	L(t,x) = -\frac{2(\alpha V - \beta)}{\alpha^2 +(\alpha V - \beta)^2z^2},\quad z=x-Vt,\\
&	\omega=-c^2/4,\\
&	\phi(z) = \frac{V}{2}z + \arctan\left(\frac{(\alpha V-\beta)z}{\alpha}\right)+\phi_0\,.
\end{align}
\end{subequations}
%By taking $\beta=0$, the solution for the Newell system reads
%\begin{subequations}
%\begin{align}\label{eq15N}
%&    S(x,t) = \frac{|V|}{|\alpha|}e^{i(\phi(z)-\omega t)}\sqrt{\frac{\alpha}{1+V^2z^2}},\\
%&	L(t,x) = -\frac{2V}{\alpha(1 + V^2z^2)},\quad z=x-Vt,\\
%&	\omega=-c^2/4,\\
%&	\phi(z) = \frac{V}{2}z + \arctan\left(Vz\right)+\phi_0,
%\end{align}
%\end{subequations}
%where $\alpha>0$, and this is also another new solution for \eqref{N}.
%
The short wave $|S|$ and long wave $L$ are bright rational solutions with amplitudes of  $\sqrt{V(\alpha V-\beta)/(\alpha^2)}$ and $-2(\alpha V-\beta)/(\alpha^2)$, respectively, on a zero background. To the best of our knowledge, this is also a novel solution of \eqref{N}.

\section{Conservation laws}
\label{sec:cl}
The YON model \eqref{YON} is integrable and therefore it allows infinitely many conservation laws. In this section, we are interested in deriving a few explicit ones, by finding convenient multipliers.
%{\color{red} check this makes sense} \textcolor{blue}{No, conservation laws are not being obtained from symmetries. There is a relation between symmetries and conservation laws, but symmetries need not to be local. Here, they have no relation at all}
A conservation law of \eqref{YON} corresponds to an expression
%to the vanishing of the divergence of a vector $\mathcal{C}=(\rho,f)$
%$$\text{Div}\,\, C = D_t C^0 + D_x C^1\Bigg|_{\text{on \eqref{YON}}}=0.$$
\begin{equation}
\label{eq:cons_law}
\rho_t + f_x=0,
\end{equation}
where $\rho\equiv \rho(S, L, S_x, L_x,\dots)$ is the density and $f\equiv f(S, L, S_x, L_x,\dots)$ the corresponding flux, respectively.
For instance, a trivial conservation law of the YON system \eqref{YON} is
\begin{subequations}
\label{eq:cons_law_1}
\begin{align}
&\rho_0= L,\\
&f_0=-2|S|^2
\end{align}
\end{subequations}
which coincides with the second equation in \eqref{YON}.

In \cite{AB1,AB2,AB3}, it was established that every conservation law corresponds to a symmetry of the equation, but very often this symmetry is non-classical or even non-local. For cases where Noether's theorem is applicable, this correspondence is explicit as Noether symmetries lead to (possibly trivial) conserved vectors; however, we ruled out this approach as we have not been able to find a variational formulation for system \eqref{YON}.
%{\color{red} This should be better explained: odd order of what?}
Instead of using Noether's approach, we will consider the direct method illustrated in \cite{AB1,AB2,AB3,Olver}, consisting in finding a vector $g=(g_1,g_2,g_3)$, called multiplier, that depends on $S$, $L$ and their derivatives up to a certain fixed but arbitrary order, such that
\begin{equation}
\frac{\delta g_1\mathcal{F}_1}{\delta S}=0,\quad \frac{\delta g_2\mathcal{F}_2}{\delta S^{\ast}}=0,\quad \frac{\delta g_3\mathcal{F}_3}{\delta L}=0,
\end{equation}
where
%\begin{align}
%\label{eqRealComplexExp}
%&\mathcal{F}_1=-S_{2,t}+S_{1,xx} +\alpha^2L^2S_1 - \beta LS_1 - 2\alpha S_1S_2^2-2\alpha S_1^3 -\alpha L_xS_2,\\
%%
%&\mathcal{F}_2=S_{1,t}+S_{2,xx}+ \alpha L_xS_1 - \beta LS_2 -2\alpha S_2^3 -2\alpha S_1^2S_2 + \alpha^2L^2S_2,\\
%%
%&\mathcal{F}_3=L_t - 2(S_1^2 + S_2^2)_x.
%\end{align}
\begin{equation}
\mathcal{F}_1=i S_t +S_{xx} +\left(i\alpha L_x+\alpha^2 L^2-\beta L -2\alpha |S|^2 \right) S\,,\quad
%\mathcal{F}_1=-S^{\ast}_t +S_{xx} -\alpha L_x S^{\ast}+\alpha^2 L^2 S-\beta L S -2\alpha |S|^2 S\,,\quad
\mathcal{F}_2=\mathcal{F}^{\ast}_1\,,\quad
\mathcal{F}_3=L_t-2(|S|^2)_x\,,
\end{equation}
and where $\delta/\delta u$ is the variational derivative with respect to the variable $u$. If such a vector $g$ can be found, then it ensures the existence of a $\rho$ and $f$ such that
\begin{equation}
\rho_t + f_x = g_1\mathcal{F}_1+g_2\mathcal{F}_2+g_3\mathcal{F}_3=0\,.
\end{equation}

By making use of GeM \cite{Chev1,Chev2,Chev3,Chev4,Chev5}, for any choices of $\alpha,\beta\in\mathbb{R}$, we can find some pairs of conserved densities and fluxes depending on derivatives up to the second order, the simples ones of which read
%, which are reported in Appendix A.
\begin{subequations}
\label{eq:cons_v_1}
\begin{align}
&\rho_1 =\frac{\alpha}{2}L^2 - |S|^2,\\
&f_1 = -2\alpha L|S|^2 - 2\, Im(S^{\ast}S_x),
%\label{energyb}
\end{align}
\end{subequations}
\begin{subequations}
\label{eq:cons_v_2}
\begin{align}
\rho_2 &= 2\alpha^2L|S|^2 -\beta|S|^2 + 2\alpha\, Im(S^{\ast}S_x),\\
f_2 &=  2\alpha |S_x|^2 - 2\alpha\, Re(S^{\ast}S_{xx}) + 4\alpha^2L\, Im(S^{\ast}S_x)-2\beta\, Im(S^{\ast} S_x),
\end{align}
\end{subequations}
while the remaining ones can be found in Appendix A.

\section{Conclusions}
%version 1
%Models describing long wave-short wave resonant interactions arise under certain assumptions in a variety of physical contexts, from fluid dynamics to plasma physics.
%We consider here a recently proposed model (see \cite{CDLS2021}) which combines the interaction terms of two integrable models, one  proposed by Yajima-Oikawa and the other one by Newell, and study some relevant families of periodic and solitary wave solutions which
%display the generation of very long waves. generalising some of the results presented in \cite{HeMeng}.
%We also derive a few conservation laws, which are of interest in view of numerical and analytical studies of this system. An argument based on the effect of the surface tension on the dispersion relation for short waves, allowing for short-long wave resonance, was presented to justify the physical relevance of the YON model in a fluid-dynamical context (and in particular for experimental set-ups of capillarity-gravity wave propagation on the scales of the centimetres), where the value of the wave number strongly depends on the water depth. In spite of the expected physical relevance, a derivation, via multiscale techniques, of the full YON system as an (integrable) reduction from a known physical PDE has not yet been achieved and remains a fascinating open problem. It is worth observing that a subcase of the YON model, namely the Yajima-Oikawa model, has been indeed derived via the multiscale technique in more than one way \cite{CDJ12000}, suggesting that this should be possible also for the more general YON model.
%version 2
Models describing long wave-short wave resonant interactions arise in a variety of physical contexts, from fluid dynamics to plasma physics. In this paper we consider the recently-proposed, long wave-short wave YON (Yajima-Oikawa-Newell) model (see \cite{CDLS2021}), an integrable model featuring two arbitrary parameters, and unifying and generalising the Yajima-Oikawa model and the Newell model.
%a recently proposed wave-short wave model (see \cite{CDLS2021}) which combines the interaction terms of two integrable models, one proposed by Yajima-Oikawa and the other one by Newell, and
%We studied some relevant families of periodic and solitary wave solutions, displaying the generation of very long waves. We also

We studied some relevant families of periodic and solitary wave solutions, displaying the generation of very long waves. Among others, we also display the expression of solutions that we term, with some abuse of language, ``rational''. Differently from the NLS equation, where the amplitude $|S|$ is indeed rational (\textit{cf.}, the Peregrine soliton), in the present case it is rather the function $|S|^2$ that comes to be rational. This is due to the quintic nonlinearity appearing in the YON system for the short wave amplitude $|S|$, rather then the usual cubic one as in the NLS equation. An analytical study of the stability of the solutions presented in this paper is left to future investigation. In this respect, we limit ourselves to report here that we carried out a preliminary numerical study, solving the initial value problem for initial conditions obtained by computing our solutions at $t=0$, using the method of lines with pseudospectral, Fourier discretisation in space and an adaptive Dormand-Prince embedded Runge-Kutta method for the time stepping: the numerical results seem to suggest the existence of regions of stability and regions where different forms of instability are observed, similarly to what is predicted for plane wave solutions of the YON system \cite{CDLS2021,DLS2018}.
%A detailed report of these numerical results requires a separated publication, and does not fit the scope of the present analytical work. 

The families of explicit solutions presented in this paper have been obtained by choosing a suitable Ansatz. A systematic derivation of soliton solutions of bright, gray and dark type, as well as of breathers and rogue-waves, exploiting the integrable character of the YON system, is currently in progress.

In this paper we have also derived a few conservation laws, which are of interest in view of numerical and analytical studies of this system. An argument based on the effect of the surface tension on the dispersion relation for short waves, allowing for short-long wave resonance, is presented to justify the physical relevance of the YON model in a fluid-dynamical context (and in particular for experimental set-ups of capillarity-gravity wave propagation on the scales of the centimetres), where the value of the wave number strongly depends on the water depth. In spite of the expected physical relevance, a derivation, via multiscale techniques, of the full YON system -- similarly to the Newell model, which is contained within the YON system -- as an (integrable) reduction from a known physical PDE has not yet been achieved and remains an intriguing open problem. It is worth observing that a subcase of the YON model, namely the Yajima-Oikawa model, has been indeed derived via the multiscale technique in more than one way \cite{CDJ12000}, suggesting that this should be possible also for the more general YON model.\\[2\baselineskip]

\textbf{Acknowledgments:} SL and PLdS acknowledge support by the Royal Society; PLdS is supported by the Royal Society under a Newton International Fellowship (reference number 201625) hosted by SL. The work of MS has been carried out under the auspices of the Italian GNFM (Gruppo Nazionale Fisica Matematica), INdAM (Istituto Nazionale di Alta Matematica).

\appendix
\section[\appendixname~\thesection]{Appendix: Conserved vectors depending on derivatives up to the second order}
%\subsection[\appendixname~\thesubsection]{}
In this Appendix we list the remaining densities and fluxes (in addition to \eqref{eq:cons_v_1} and \eqref{eq:cons_v_2}) depending on derivatives up to the second order.%, along with the special cases for $\alpha=0$ and $\beta=0$.
 These have been computed using GeM \cite{Chev1,Chev2,Chev3,Chev4,Chev5}, for $\alpha,\beta\in\mathbb{R}$.

%%%%
%
%\begin{subequations}
%\begin{align}
%&\rho = L|S|^2 -\frac{\beta}{2\alpha^2}|S|^2 + \frac{2}{\alpha}S_1S_{2,x},\\
%&f = \alpha L^2|S|^2 -\frac{\beta}{\alpha}L|S|^2- 2|S|^4 + \frac{1}{\alpha}|S_x|^2 - \frac{2}{\alpha}S_1S_{2,t} + 2L(S_1S_{2,x}-S_{1,x}S_2)-\frac{\beta}{\alpha}(S_1S_{2,x} - S_{1,x}S_2);
%\end{align}
%\end{subequations}
%

\begin{subequations}
\begin{align}
\rho_3 =& 2\alpha^2 tL|S|^2+\frac{\alpha^2}{2}xL^2-\frac{\beta}{2} xL-\beta t|S|^2-\alpha x|S|^2+2\alpha t\, Im(S^{\ast}S_x),\\
f_3 =& - 2\alpha^2 xL|S|^2 + \beta x|S|^2+ 4\alpha^2tL\, Im(S^{\ast}S_x) +2\alpha t|S_x|^2-2\alpha x\, Im(S^{\ast}S_x)\\
&-2\beta t\, Im(S^{\ast}S_x) - 2\alpha t\, Re(S^{\ast}S_{xx});\notag
%
%
%\rho =& tL|S|^2+\frac{1}{4}xL^2-\frac{\beta}{4\alpha^2}xL-\frac{\beta}{2\alpha^2}t|S|^2-%\frac{1}%{2\alpha}x|S|^2+\frac{2}{\alpha}tS_1S_{2,x},
%
%f =& \frac{1}{2\alpha^2}\left(2\alpha^3 tL^2|S|^2 - 4 \alpha^2t|S|^4 - 2\alpha^2 xL|S|^2 - %2\alpha\beta tL|S|^2+ \beta x|S|^2\right.+ 4\alpha^2tL(S_1S_{2,x}-S_{1,x}S_2) +2\alpha t|S_x|^2\\
%&\left.-2\alpha x(S_1S_{2,x}-S_{1,x}S_2)-2\beta t(S_1S_{2,x}-S_{1,x}S_2)-2\alpha S_1S_2 - 4\alpha t %S_1S_{2,t}\right);\notag
\end{align}
\end{subequations}

\begin{subequations}
\begin{align}
\rho_4 =&\frac{1}{2}\alpha^5L^4 -2\alpha^4 L^2|S|^2 -\alpha^3\beta L^3 +2\alpha^3|S|^4+4\alpha^3L\, Im(S^{\ast}S_x)-\frac{1}{2}\alpha^3L_x^2 +2\beta^2 |S|^2-4\alpha\beta\, Im(S^{\ast}S_x) +4\alpha^2|S_x|^2,\\
%
%
%\frac{1}{8\alpha^3}\left(\alpha^5L^4 -4\alpha^4L^2(S_1^2+S_2^2) -2\alpha^3\beta L^3 +4\alpha^3(S_1^2+S_2^2)^2+8\alpha^3L(S_1S_{2,x} -S_{1,x}S_2)\right.\notag\\
%&\left.-\alpha^3L_x^2 +4\beta^2(S_1^2+S_2^2) -16\alpha\beta S_1S_{2,x} +8\alpha^2(S_{1,x}^2 +S_{2,x}^2)\right),\\
%
%
%f_4 =& -\frac{1}{4}\left(-2\alpha^3\beta L^2|S|^2 +4\alpha^4 L^2\, Im(S^{\ast}S_x) +4\alpha^2\beta|S|^4 %-4\alpha^3L|S|^2 -2\alpha^3L_x(|S|^2)_x + 4\alpha\beta|S_{x}|^2+\right.\\
%&-24\alpha^3|S|^2\, Im(S^{\ast}S_x) + 4\alpha^3L\, Re(S^{\ast}S_{xx})+4\alpha^5L^2|S|^2-4\alpha^3\beta L^2|S|^2- %8\alpha^4L|S|^4 + 8\alpha^2\, Re(S^{\ast}_xS_{xx})-4\alpha^3L_x\, Re(S^{\ast}S_x)+\notag\\
%&\left.-4\beta^2\, Im(S^{\ast}S_x)-4\alpha\beta\, Re(S^{\ast}S_{xx})\right)\notag.
%
f_4 =&6\alpha^3\beta L^2|S|^2+8\alpha^4 L |S|^4 -4\alpha^2 \beta |S|^4 - 4\alpha \beta |S_x|^2+4\alpha^3 L|S_x|^2-4\alpha^5 L^3|S|^2+8\alpha^3 |S|^2\, Im(S^{\ast}S_x)\\
&+4\beta^2\, Im(S^{\ast}S_x)-4\alpha^4 L^2\, Im(S^{\ast}S_x) + 8 \alpha^3 L_x\, Re(S^{\ast}S_x)+4\alpha\beta\, Re(S^{\ast}S_{xx})-4\alpha^3 L\, Re(S^{\ast}S_{xx})+8\alpha^2\, Im(S^{\ast}_xS_{xx})\notag.
\end{align}
\end{subequations}


\begin{thebibliography}{99}

\bibitem{Whitham} Whitham, G. B., Nonlinear dispersion of water waves, {\em J. Fluid Mech.}, {\bf 1966}, {\em 27},  399--412.

\bibitem{AKNS} Ablowitz, M. J., Kaup, D. J., Newell, A. C., Segur, H., The inverse scattering transform-Fourier analysis for nonlinear problems, {\em Stud. Appl. Math.}, {\bf 1974}, {\em 53}(4), 249--315.

\bibitem{CalDe} Calogero, F., Degasperis, A., {\em Spectral Transform and Solitons}. Amsterdam: North-Holland, {\bf 1982}.

\bibitem{Zakharov-Kuznetsov1986} Zakharov, V. E., Kuznetsov, E. A., Multi-scale expansions in the theory of systems integrable by the inverse scattering transform, {\em Physica D}, {\bf 1986}, {\em 18D},  455--463.

\bibitem{Calogero1991} Calogero, F., Why are certain nonlinear PDEs both widely applicable and integrable?, in {\em What is integrability?} (ed. V. E. Zakharov). Berlin: Springer-Verlag, {\bf 1991}.

\bibitem{Degasperis2009} Degasperis, A., Multiscale expansion and integrability of dispersive wave equations, in {\em Integrability}, vol. 767 (ed. A. V. Mikhailov). Lecture Notes in Physics. Berlin: Springer-Verlag, {\bf 2009},  215--244.

\bibitem{Benney1977} Benney, D. J., A general theory for interactions between short and long waves, {\em Stud. Appl. Math.}, {\bf 1976}, {\em 56}(1), 81--94.

\bibitem{YO1976} Yajima, N., Oikawa, M., Formation and interaction of sonic-Langmuir solitons: Inverse scattering method, {\em Prog. Theor. Phys.}, {\bf 1976}, {\em 56}, 1719--1739.

\bibitem{CDJ12000} % not sure
Calogero, F., Degasperis, A., Xiaoda, J., Nonlinear Schr\"{o}dinger-type equations from multiscale reduction of PDEs. I. Systematic derivation, {\em J. Math. Phys.}, {\bf 2000}, {\em 41}, 6399--6443.

\bibitem{Newell1978} Newell, A. C., Long waves-short waves: A solvable model, {\em SIAM J. Appl. Math.}, {\bf 1978}, {\em 35}, 650--664.

\bibitem{CDLS2021} Caso-Huerta, M., Degasperis, A., Lombardo, S.,Sommacal, M., A new integrable model of long wave-short wave interaction and linear stability spectra, {\em Proc. R. Soc. A.}, {\bf 2021}, {\em 477}, 20210408.

\bibitem{Wright2006} Wright, O. C., Homoclinic connections of unstable plane waves of the long-wave--short-wave equations, {\em Stud. Appl. Math.}, {\bf 2006}, {\em 117}, 71--93 [in particular, see system (53)].

\bibitem{CCFMO2018}
Chen, J., Chen, Y., Feng, B-F., Maruno, K., Ohta, Y., General high-order rogue waves of the (1+ 1)-dimensional Yajima-Oikawa system,  {\em J. Phys. Soc. Jpn.},  {\bf 2018}, {\em 87}(9), 094007.

\bibitem{CG2022}
Li, R., Geng, X., Periodic-background solutions for the Yajima-Oikawa long-wave--short-wave equation, {\em Nonlinear Dyn.}, {\bf 2022}, {\em 94}.

\bibitem{CT2008} Chowdhury, A., Tataronis, J. A., Long-wave short-wave resonance in nonlinear negative refractive index media, {\em Phys. Rev. Lett.}, {\bf 2008}, {\em 100}(15), 153905.

\bibitem{DR1977} Djordjevic, V. D., Redekopp, L. G., On two-dimensional packets of capillary-gravity waves, {\em J. Fluid Mech.}, {\bf 1977}, {\em 79}(4), 703--714.

\bibitem{Lannes2013} Lannes, D., {\em The Water Waves Problem: Mathematical Analysis and Asymptotics}. Mathematical Surveys and Monographs, vol. 188. Providence: American Mathematical Society, {\bf 2013}.

\bibitem{Grimshaw1977} Grimshaw, R. H. J., The modulation of an internal gravity-wave packet, and the resonance with the mean motion, {\em Stud. Appl. Math.}, {\bf 1977}, {\em 56}, 241--266.

\bibitem{KR1981} Koop, C. G. and Redekopp, L. G., The interaction of long and short internal gravity waves: Theory and experiment, {\em J. Fluid. Mech.} , {\bf 1981}, 367--409.

\bibitem{AB1} Anco, S. C., Bluman, G. W., Direct construction of conservation laws from field equations, {\em Phys. Rev. Lett.}, {\bf 1997}, {\em 78}, 2869--2873.

\bibitem{AB2} Anco, S. C., Bluman, G. W., Direct construction method for conservation laws of partial differential equations. I: Examples of conservation law classifications, {\em Eur. J. Appl. Math.}, {\bf 2002}, {\em 13}, 545--566.

\bibitem{AB3} Anco, S. C., Bluman, G. W.,  Direct construction method for conservation laws of partial differential equations. II: General treatment,  {\em Eur. J. Appl. Math.}, {\bf 2002}, {\em 13},  567--585.

\bibitem{Chev1} Cheviakov, A., GeM software package for computation of symmetries and conservation laws
of differential equations, {\em Comput. Phys. Commun.},  {\bf 2007}, {\em 176}, 48--61.

\bibitem{Chev2} Cheviakov, A.,  Symbolic computation of local symmetries of nonlinear and linear partial and ordinary differential equations, {\em Math. Comput. Sci.}, {\bf 2010}, {\em 4}, 203--222.

\bibitem{Chev3} Cheviakov, A., Computation of fluxes of conservation laws, {\em J. Eng. Math.}, {\bf 2010}, {\em  66}, 153--173.

\bibitem{Chev4} Cheviakov, A., Symbolic computation of nonlocal symmetries and nonlocal conservation laws of partial differential equations using the GeM package for Maple, in {\em Similarity and Symmetry Methods} (eds J.-F. Ganghoffer, I. Mladenov). Lecture Notes in Applied and Computational Mechanics, vol. 73. Cham: Springer, {\bf 2014}, 165--184.

\bibitem{Chev5} Cheviakov, A., Symbolic computation of equivalence transformations and parameter reduction
for nonlinear physical models, {\em Comput. Phys. Commun.}, {\bf 2017}, {\em 220}, 56--73.

\bibitem{Olver} Olver, P. J., {\em Applications of Lie Groups to Differential Equations}, 2nd edition. Graduate Texts in Mathematics, vol. 107. New York: Springer, {\bf 1993}.
    
\bibitem{DLS2018} Degasperis, A., Lombardo, S., Sommacal, M., Integrability and Linear Stability of Nonlinear Waves, {\em J. Nonlinear Sci.}, {\bf 2018}, {\em 28(4)}, 1251--1291.

\end{thebibliography}
\end{document}